\UseRawInputEncoding
\documentclass[10pt,journal,compsoc]{IEEEtran}

\ifCLASSOPTIONcompsoc
  \usepackage[nocompress]{cite}
\else
  \usepackage{cite}
\fi
\usepackage{multirow}
\usepackage{makecell}
\usepackage{amsmath}
\usepackage{amsthm}
\usepackage{amssymb}
\usepackage{graphicx}
\usepackage{algpseudocode}
\usepackage{algorithm, algorithmicx}
\usepackage{booktabs}
\usepackage{threeparttable}
\newtheorem{definition}{Definition}
\usepackage{enumitem}

\ifCLASSINFOpdf

\else

\fi

\begin{document}

\title{Petri Net Based Symbolic Model Checking for Computation Tree Logic of Knowledge}

\author{Leifeng He and Guanjun Liu

\IEEEcompsocitemizethanks{\IEEEcompsocthanksitem L. F. He and G. J. Liu are with the Department of Computer Science, Tongji University, Shanghai 201804, China.\protect\\

E-mail: liuguanjun@tongji.edu.cn
}}

\IEEEtitleabstractindextext{%
\begin{abstract}
Computation Tree Logic of Knowledge (CTLK) can specify many design requirements of privacy and security of multi-agent systems (MAS). In our conference paper, we defined Knowledge-oriented Petri Nets (KPN) to model MAS and proposed Reachability Graphs with Equivalence Relations (RGER) to verify CTLK. In this paper, we use the technique of Ordered Binary Decision Diagrams (OBDD) to encode RGER in order to alleviate the state explosion problem and enhance the verification efficiency. We propose a heuristic method to order those variables in OBDD, which can well improve the time and space performance of producing, encoding and exploring a huge state space. More importantly, our method does not produce and encode any transition or equivalence relation of states when producing and encoding an RGER, and in fact it dynamically produces those transition or equivalence relations that are required in the verification process of CTLK formulas. This policy can save a lot of time and space since the number of transition or equivalence relations of states is much greater than the number of states themselves. We design symbolic model checking algorithms, develop a tool and apply them to two famous examples: Alice-Bob Protocol and Dining Cryptographers Protocol. We compare our tool with MCMAS which is the state-of-the-art model checker of verifying CTLK. The experimental results illustrate the advantages of our model and method. Our tool running in a general PC can totally spend less than 14 hours to verify Dining Cryptographers Protocol with 1200 concurrent cryptographers where there are about $10^{1080}$ states and the two verified CTLK formulas have more than 6000 atomic propositions and more than 3600 operators. These good performances are owed to a combination of the OBDD technique and the structure characteristics of KPN.
\end{abstract}

}

\maketitle

\IEEEdisplaynontitleabstractindextext

\IEEEpeerreviewmaketitle

\IEEEraisesectionheading{\section{Introduction}\label{sec:introduction}}
Errors in some privacy/security-critical systems such as anonymity protocols can lead to serious consequences. It is therefore important for designers to use some logically precise approaches to find these errors. \emph{Model checking}~\cite{MC2,MC3,MC1,MW} is an automated and practically successful approach of formally verifying these systems.

In the paradigm of model checking, a system is first encoded by a model. Famous modelling languages include \emph{labelled transition systems} (LTS)~\cite{MC2}, Petri nets~\cite{PNMC1}, \cite{PNMC3},~\cite{PNMC2},~\cite{PNMC4}, reactive modules~\cite{REA}, \emph{New Symbolic Model Verifier} (NUSMV)~\cite{MCTK,NUSMV}, \emph{Interpreted Systems Programming Language} (ISPL)~\cite{MCMAS} and so on. Then, a to-be-checked property (or requirement) of the system is specified by a logical formula. Verifying whether the system satisfies the property is translated into the problem of checking whether the model satisfies the logical formula. Some design requirements, such as deadlock-freeness, safety, liveness and fairness, can be specified by discrete temporal logic such as \emph{Linear Temporal Logic} (LTL)~\cite{MC2,LTL2,LTL3} and \emph{Computation Tree Logic} (CTL)~\cite{CTL2},~\cite{CTL},~\cite{PNMC2}.

A big challenge in model checking is the \emph{state explosion problem} especially for concurrent systems, i.e., the state space of a system grows exponentially with the number of variables. Many techniques have been developed to deal with this problem such as OBDD-based symbolic model checking~\cite{OBDD-SMC1,MC2,sy-BPN,MCMAS,MARCIE}, SAT-based bounded model checking~\cite{BMC,CTLK,SAT}, abstraction~\cite{ABS1,ABS2,MW2}, partial order reduction~\cite{RED,JUNSUN} and decomposition~\cite{DECOM}.

A \emph{Multi-Agent System} (MAS)~\cite{MAS2,MAS3,MAS4,MAS5} is a concurrent system where multiple agents interact or collaborate with each other in order to perform some common/distributed tasks. Both the correctness of interacting/collaborating behaviors and the privacy/security of agents should be ensured before an MAS is put into service. Discrete temporal logic can only specify some design requirements of interacting/collaborating behaviors, but cannot specify the requirements of privacy and security. Then \emph{Epistemic logic}~\cite{EL,EL2} (or \emph{logic of knowledge}) is used to specify these privacy/security-related requirements since they can represent knowledge of agents.

Epistemic logic was first put forward by philosophers and later used in the computer science field. Nowadays it has become a means of reasoning about the knowledge and belief of agents~\cite{BTP},~\cite{MCMAS},~\cite{CTLK}. It is a modal logic concerned with agent-related reasoning and offers a useful expression and analysis of privacy and security of MAS. It has been used for verifying \emph{security protocols}~\cite{SEP}, \emph{agreement protocols}~\cite{AGP} and some other knowledge-related systems~\cite{MCMAS}. By adopting epistemic modalities as primitives, one can naturally express private and collective (common or distributed) knowledge of agents. If a fact is private to some agents, then only these agents know it but others do not know it at all. As a kind of epistemic logic, \emph{Computation Tree Logic of Knowledge} (CTLK)~\cite{CTLK} is a temporal and epistemic logic\footnote{If a CTLK formula has no epistemic operator, then it is a CTL formula in fact.} and can specify many design requirements of MAS including not only privacy/security behaviors but also interacting/collaborating behaviors. \emph{Bit Transmission Protocol}~\cite{BTP,ToforMC2} and \emph{Dining Cryptographers Protocol}~\cite{Din3,Din2,Din1} can both be viewed as an MAS and their privacy/security-related requirements can be specified by CTLK. Some algorithms and tools have been developed for the related model checking~\cite{MCMAS,AlforMC}.

MCMAS~\cite{MCMAS} is a state-of-the-art model checker of verifying CTLK, where an MAS is modelled by ISPL and the requirements of privacy/security are specified by CTLK. However, it usually has three weaknesses:
\begin{itemize}
\item[1.] Programs written by ISPL are usually non-intuitive and have hard readability, which was also pointed out in~\cite{MCTK}. Additionally, a program of ISPL needs an environment agent to configure the interaction/collaboration of all agents, which usually brings much inconvenience to users when building or scaling up a model.
\item[2.] It has a high time complexity when translating a program of ISPL into its induced model as its behavioral representation. Although it uses OBDD to alleviate the state explosion problem, the time complexity of producing an induced model is still very high since it needs to find a good compromise between continuous variable reordering in OBDD and efficiency of reducing memory-consuming.
\item[3.] Before verifying a CTLK formula, it first needs to produce all states as well as all transition and equivalence relations of these states. As we all know, the number of transition or equivalence relations of states is much greater than the number of states themselves. But in fact, it is not necessary to produce all transition and equivalence relations of states when verifying a CTLK formula. Thus this also wastes too much time.
\end{itemize}

In order to overcome these weaknesses, this paper uses Petri nets and OBDD to model MAS and to verify CTLK since we can make full use of the structure characteristics of Petri nets. There are some studies on Petri-nets-based model checking for MAS~\cite{PNMC1,PNMC2,PNMC4,PNMC3}. However, they only pay attention to the correctness of interacting/collaborating behaviors but do not consider the privacy/security-related requirements. Therefore, in our conference paper~\cite{KPN}, we defined \emph{Knowledge-oriented Petri Nets} (KPN) to intuitively simulate both the processes of interaction/collaboration of multiple agents and their epistemic evolutions. We used CTLK~\cite{CTLK} to specify the requirements of both interaction/collaboration and privacy/security. We defined equivalence relation for each agent and constructed \emph{Reachability Graph with Equivalence Relations} (RGER)\footnote{In~\cite{KPN}, we called it as \emph{similar reachability graph}. Since a similar relation defined in~\cite{KPN} is an equivalence relation, we rename it as \emph{Reachability Graph with Equivalence Relations} in this paper.} to verify CTLK. We designed their model checking algorithms, but the state explosion problem was not handled well. In this paper, we consider more epistemic operators in CTLK, improve our algorithms greatly and develop a related tool. We use OBDD~\cite{BDD,CUDD} as the symbolic representation of states of RGER in order to alleviate the state explosion problem. Especially, we propose a heuristic method to order variables in OBDD that can well improve the time and space performance of producing, encoding and exploring the states of an RGER. The heuristic method exactly utilizes the structure characteristics of KPN, but that is not easy for MCMAS since its modeling language ISPL does not have an obvious or direct structure representation. More importantly, instead of producing all transition and equivalence relations of states before verification, our algorithms dynamically produce those transition or equivalence relations that are required in the process of verifying a set of CTLK formulas. This also makes full use of the structure characteristics of KPN. Therefore, our algorithms can save much time. Our experiments over the benchmark of Dining Cryptographers Protocol show that our tool can obtain surprising results compared with MCMAS. We also compare our heuristic method of ordering variables in OBDD with the state-of-the-art one proposed in~\cite{heu} and show that our method can obtain a good overall performance.

The remainder of this paper is organized as follows. Section~2 introduces some basic concepts of Petri nets. Section~3 introduces KPN and RGER. Section~4 recalls OBDD and presents an OBDD-based symbolic approach to produce and encode the states and relations of an RGER. Section~5 introduces the syntax and semantics of CTLK based on KPN and RGER. Section~6 proposes our model checking algorithms in which we use the structure of KPN and the encoded states of RGER to produce those required transition or equivalence relations. Section~6 also shows our tool briefly. Section~7 uses two examples to show the usefulness of our model and method. Section~8 illustrates and analyzes our experiments. Section~9 concludes this paper.

\section{Petri Nets}
Petri nets and their related concepts are recalled in this section. For more details, one may refer to~\cite{PN1} and~\cite{PN2}. $\mathbb{N}=\{0$, $1$, $2$, $\cdots\}$ is the set of all non-negative integers.

A \emph{net} is a 3-tuple $N=(P$, $T$, $F)$ where $P=\{p_1$, $\cdots$, $p_n\}$ is a finite set of \emph{places}, $T=\{t_1$, $\cdots$, $t_m\}$ is a finite set of \emph{transitions}, $F\subseteq(P\times T)\cup (T\times P)$ a set of \emph{arcs}, and $P\,\cap\, T=\emptyset$. A net can be viewed as a directed bipartite diagram. Generally, transitions are represented by rectangles and places by circles in a net diagram. Given a net $N=(P$, $T$, $F)$ and a node $x\in P$ $\cup$ $T$, the \emph{pre-set} and \emph{post-set} of $x$ are defined as $^\bullet x=\{y\in P$ $\cup$ $T\mid (y$, $x)\in F\}$ and $x^\bullet =\{y\in P$ $\cup$ $T\mid (x$, $y)\in F\}$, respectively. If the post-set of a place is empty, we call it as an \emph{end place}.

A \emph{marking} of $N=(P$, $T$, $F)$ is a mapping $M$: $P\rightarrow \mathbb{N}$ where $M(p)$ is the number of tokens in place $p$. Place $p$ is \emph{marked} at $M$ if $M(p)>0$. A marking is denoted as a multi-set of places in this paper. For example, if the marking $M$ satisfies that place $p_1$ has 2 token, place $p_2$ has 4 tokens and other places have no tokens, then it is written as $M=\{2p_1$, $4p_2\}$. Sometimes, a marking is also called as a \emph{state} in this paper.

A net $N$ with an \emph{initial marking} $M_0$ is called a \emph{Petri net} or \emph{net system} and denoted as $(N$, $M_0)$. Transition $t$ is \emph{enabled} at $M$ if $\forall p\in\hspace{+0.1mm}^\bullet\hspace{-0.1mm} t$: $M(p)>0$. This is denoted as $M[t\rangle$. \emph{Firing} an enabled transition $t$ yields a new marking $M^\prime$ which satisfies that $M^\prime(p)=M(p)-1$ if $p\in\, ^\bullet$$ t\setminus t^\bullet$; $M^\prime(p)=M(p)+1$ if $p\in t^\bullet\setminus ^\bullet$$ t$; and $M^\prime(p)=M(p)$ otherwise. This is denoted as $M[t\rangle M^\prime$, and we call $M$ as a \emph{predecessor} of $M^\prime$. A marking $M_k$ is \emph{reachable} from another marking $M$ if $M_k=M$ or there exists a non-empty transition sequence $\sigma =t_1 t_2 \cdots t_k$ such that $M[t_1\rangle M_1[t_2\rangle\cdots \rangle M_{k-1}[t_k\rangle M_k$. We use $M[\sigma\rangle M_k$ to represent that $M$ reaches $M_k$ after firing $\sigma$. The set of all markings reachable from $M$ in a net $N$ is denoted as $R(N$,~$M)$.

The \emph{reachability graph} of a Petri net $(N$, $M_0)=(P$, $T$, $F$, $M_0)$ is a 3-tuple $\Delta = (\mathbb{M}$, $T$, $\mathbb{F})$ where $\mathbb{M} = R(N$, $M_0)$ is the set of all reachable markings and $\mathbb{F}\subseteq \mathbb{M}\times T\times \mathbb{M}$ is the set of all directed edges such that $(M$, $t$, $M^\prime)\in \mathbb{F}$ iff $M[t\rangle M^\prime$.

A Petri net is \emph{safe} if each place has at most one token in each reachable marking. In this paper, we only consider safe Petri nets and thus a marked place $p$ means $M(p)=1$. A marking $M$ is a \emph{deadlock} if every transition is disabled at $M$.

\section{KPN and RGER}
In this section, we introduce \emph{Knowledge-oriented Petri nets} (KPN) and their \emph{Reachability Graphs with Equivalence Relations} (RGER). For more details, one may refer to~\cite{KPN}.

\begin{figure} [tt]
\centering
\includegraphics[width=0.485\textwidth]{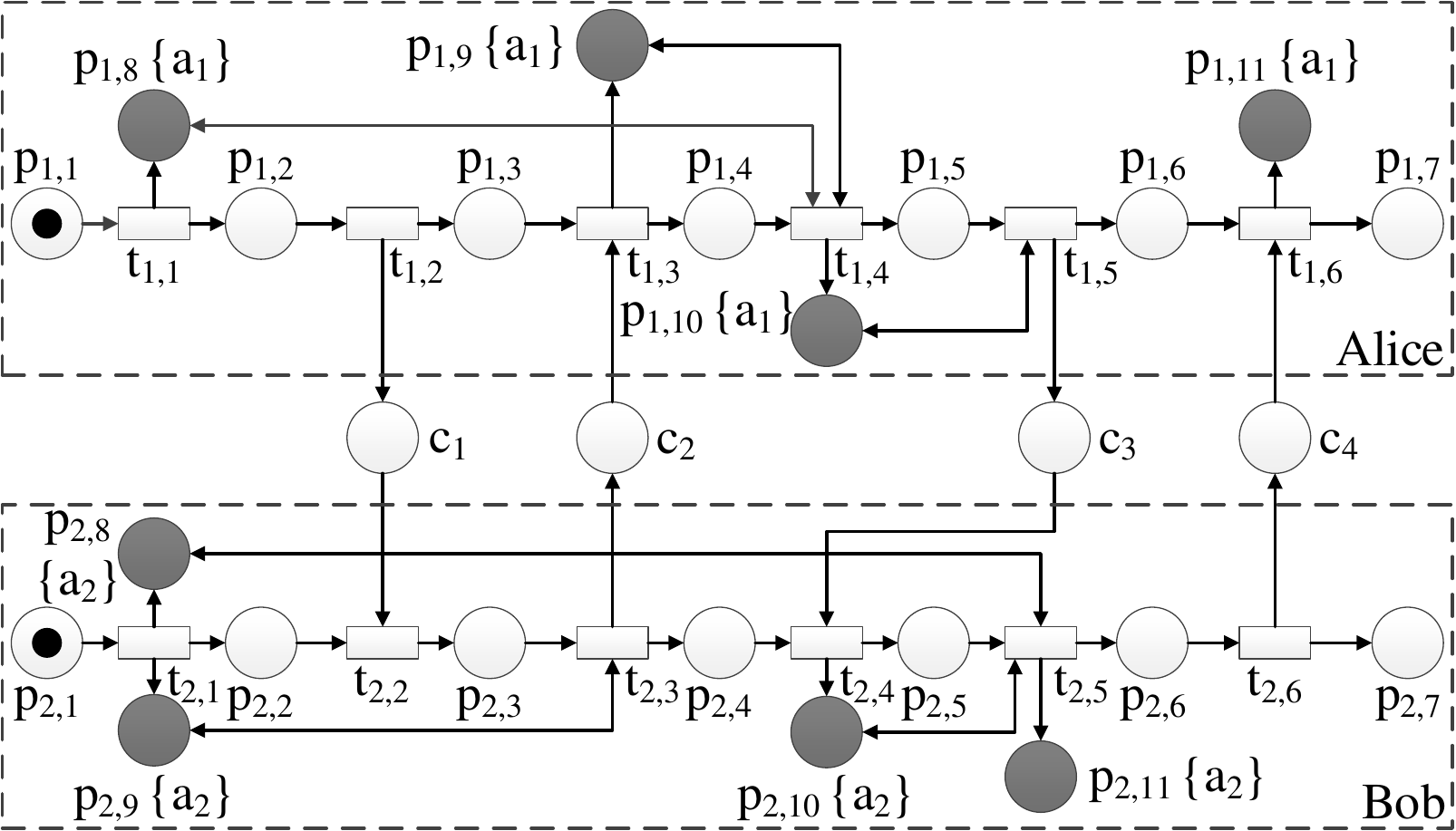}
\caption {KPN modeling Alice-Bob Protocol.}
\label{ABP}
\end{figure}
\begin{figure*}[tt]
\centering
\includegraphics[width=\textwidth]{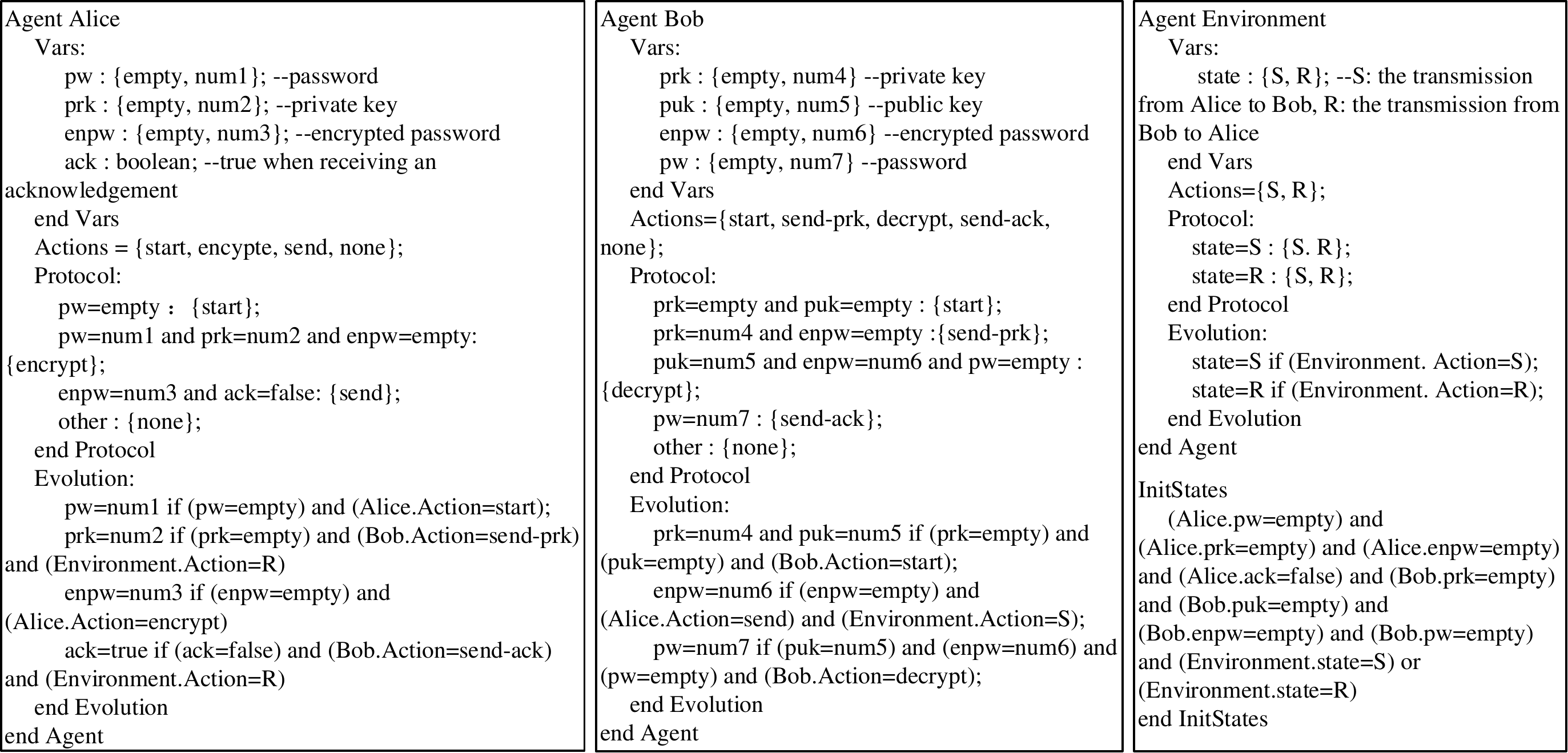}
\caption {The program of ISPL describing Alice-Bob Protocol.}
\label{ISPL}
\end{figure*}

\subsection{Knowledge-oriented Petri Nets}
\begin{definition}[KPN]
A KPN is a 7-tuple $\Sigma = (P_S$, $P_K$, $T$, $F$, $M_0$, $\mathcal{A}$, $L)$ where
\begin{itemize}
\item[1.] $(P_S\cup P_K$, $T$, $F$, $M_0)$ is a safe Petri net;
\item[2.] $P_S$ is a set of local state places\footnote{In our conference paper \cite{KPN}, we called these places as \emph{state places}. In fact, a place in a Petri net represents a local state of the related system \cite{PN1,PN2}, and a distribution of tokens across all places (i.e., a marking of $P_S\cup P_K$) represents a global state. For readability, this paper renames these places as \emph{local state places} instead of \emph{state places} in \cite{KPN}.}, $P_K$ is a set of basic knowledge places, and $P_S\cap P_K = \emptyset$;
\item[3.] $\mathcal{A} = \{a_1$, $a_2$, $\cdots$, $a_m \}$ is a set of names of agents;
\item[4.] $L:P_K\rightarrow 2^\mathcal{A}\setminus\{\emptyset\}$ is a labeling function.
\end{itemize}
\end{definition}

A KPN is a special Petri net where the epistemic evolution of each agent is considered. The underlying Petri net w.r.t. $P_S$ models the execution process of each agent and the interaction/collaboration of multiple agents. For each $p\in P_K$, it represents a basic knowledge obtained by a set of agents when it is marked, and $L(p)$ means those agents who obtain the knowledge. A place is also used to represent an atomic proposition in CTLK, i.e., an atomic proposition is true iff the corresponding place is marked, which is also the reason why a KPN is required to be safe. Therefore, KPN can model both the interaction/collaboration of multiple agents and their epistemic evolutions.

In our previous work~\cite{KPN} and other related work~\cite{MCK,MCTK,MCMAS}, some notions and methods were usually illustrated via \emph{bit transmission protocol}~\cite{BTP,ToforMC2}. Here, we use \emph{Alice-Bob Protocol}~\cite{ABP} to illustrate our notions and methods since it is closely related to privacy/security. In this protocol, a password needs to be transferred between Alice and Bob secretly and it is realized by the asymmetric encryption technique.\footnote{Here we focus on the epistemic and interacting processes of this protocol, but do not concern the specific encryption and decryption technique. Even from this perspective, we use our methods to show that this protocol is not secure.} The KPN in Fig.~\ref{ABP} models this protocol where the hollow circles are local state places and the solid circles are basic knowledge places. Here, we use $a_1$ and $a_2$ to represent the names of Alice and Bob, respectively. First of all, Alice chooses her password and Bob chooses his public and private keys, which are modelled by transitions $t_{1,1}$ and $t_{2,1}$, respectively. Secondly, Alice sends Bob a request ($t_{1,2}$) asking for his public key, and Bob delivers his public key to Alice ($t_{2,3}$) once receiving the request ($t_{2,2}$). Thirdly, after Alice receives the public key ($t_{1,3}$), she uses this key to encrypt her password ($t_{1,4}$) and sends it to Bob ($t_{1,5}$). Fourthly, after Bob receives the encrypted password ($t_{2,4}$), he uses his private key to decrypt it ($t_{2,5}$). Finally, Bob sends an acknowledgement to Alice ($t_{2,6}$).

Places $p_{1,8}$--$p_{1,11}$ and $p_{2,8}$--$p_{2,11}$ are basic knowledge places. A token in $p_{1,8}$ means that Alice owns the basic knowledge that she has got her password. A token in $p_{2,8}$ (resp. $p_{2,9}$) means that Bob owns the basic knowledge that he has got his private (resp. public) key. A token in $p_{1,9}$ means that Alice owns the basic knowledge that she has got a public key. A token in $p_{1,10}$ (resp. $p_{2,10}$) means that Alice (resp. Bob) owns the basic knowledge that she (resp. he) has got an encrypted password. A token in $p_{2,11}$ means that Bob owns the basic knowledge that he has got a password. A token in $p_{1,11}$ means that Alice owns the basic knowledge that she has got an acknowledgement. For complex knowledges, we use CTLK to specify them and verify their validity via reasoning. For example, when Alice receives an acknowledgement, she can derive that Bob has got the password.

Note that for the simplification of a KPN diagram, a self-loop is represented by an arc with arrowheads at both ends, e.g., the self-loop between $p_{1,8}$ and $t_{1,4}$ in Fig.~\ref{ABP}. A self-loop between a transition and a place means that the transition is an output of the place and the place is also an output of the transition. Additionally, in the above example, every knowledge is owned by only one agent. The case of common knowledge, i.e., a knowledge is owned by two or more agents, can be found in the example of Dining Cryptographers Protocol. Fig.~\ref{ISPL} shows the program of Alice-Bob Protocol written in ISPL. Obviously, KPN is more intuitive and has better readability.

In a KPN, each agent corresponds to a subnet and different agents interact or collaborate via some common places such as $c_1$--$c_4$ in Fig.~\ref{ABP}. Therefore, it avoids an environment agent to describe the interaction and collaboration of agents. However, MCMAS needs such an agent because it is used to controls two kind of common variables: global variables observable by all agents and local variables observable by two or more agents. In other words, the environment agent is in charge of the interaction and collaboration of all agents. Therefore, KPN has better scalability and can be expanded more easily compared to ISPL of MCMAS, especially when we do experiments facing hundreds of or even more than a thousand agents. Of course, the environment agent in Fig.~\ref{ISPL} is simple because there is no common variable. Later, we will use the example of Dining Cryptographers Protocol to show the complexity of one environment agent.

The rules of enabling and firing transitions of KPN are the same as those of Petri nets described in Section~2. Therefore, we can produce the reachability graph for any KPN. In order to reflect the epistemic evolution of each agent (e.g., whether an agent has the same knowledge at two different markings?), we define an equivalence relation for each agent based on the reachability graph and thus we can verify CTLK using them.

\begin{figure*}[tt]
\centering
\includegraphics[width=\textwidth]{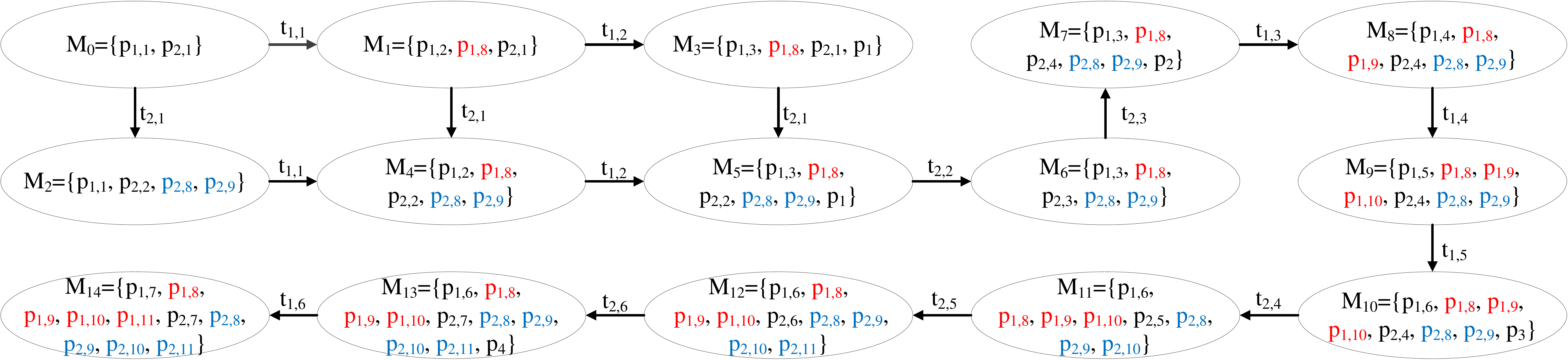}
\caption {The reachability graph of the KPN in Fig. \ref{ABP}.}
\label{ABPG}
\end{figure*}
\begin{figure*}[tt]
\centering
\includegraphics[width=\textwidth]{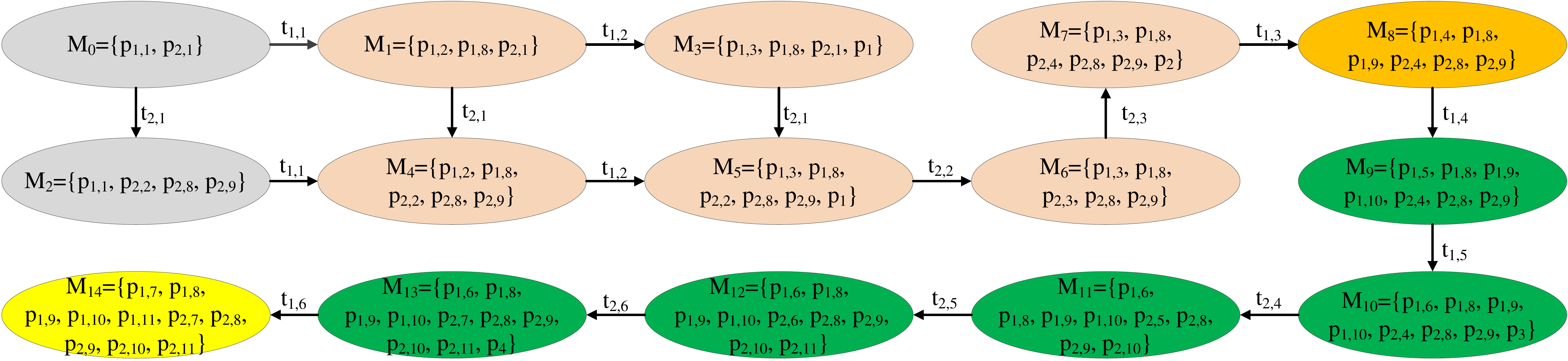}
\caption {Equivalence relation $\sim_{a_1}$ in Fig. \ref{ABPG}.}
\label{ABPG1}
\end{figure*}
\begin{figure*}[tt]
\centering
\includegraphics[width=\textwidth]{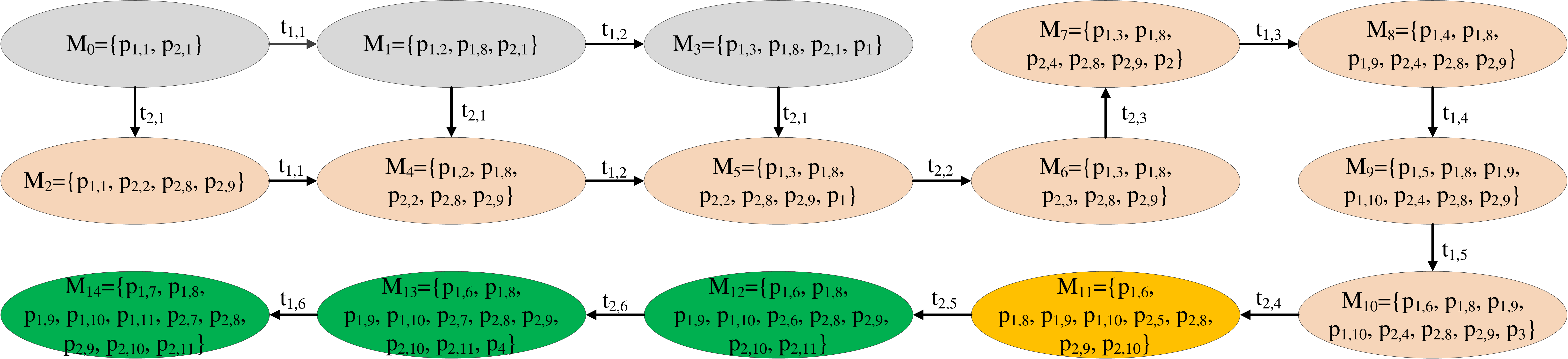}
\caption {Equivalence relation $\sim_{a_2}$ in Fig. \ref{ABPG}.}
\label{ABPG2}
\end{figure*}

\subsection{Reachability Graph with Equivalence Relations}
Given a marking $M$ and a set of places $P$ in a KPN, we use $M\upharpoonright P$ to denote the projection of $M$ onto $P$, i.e., $M\upharpoonright P = \{ p\in P\mid M(p)>0\}$. For each agent $a\in\mathcal{A}$, we use $P_a$ to represent those basic knowledge places w.r.t. agent $a$, i.e., $P_a = \{p\in P_K\mid a\in L(p)\}$. Then $M\upharpoonright P_a=\{p\in P_a\mid M(p)>0\}$ denotes the basic knowledges owned by agent $a$ at $M$. For a set of agents $\Gamma\subseteq\mathcal{A}$, we use $P_{\Gamma}$ to represent those basic knowledeg places w.r.t. $\Gamma$, i.e., $P_{\Gamma} = \{p\in P_K\mid\exists a\in\Gamma: a\in L(p)\}$. Then $M\upharpoonright P_{\Gamma}=\{p\in P_\Gamma\mid M(p)>0\}$ denotes the basic knowledges owned by at least one agent in $\Gamma$ at $M$. For example in~Fig.\ref{ABP}, $M = \{p_{1,2}$, $p_{1,8}$, $p_{2,1}\}$ is a reachable marking from $M_0$ by firing $t_{1,1}$, and all basic knowledges at $M$ are $M\upharpoonright P_{\mathcal{A}} = \{p_{1,8}\}$ which means that Alice has got a password but Bob has no any basic knowledge at $M$ (i.e., $P_{a_2} = \{p_{2,8}$, $p_{2,9}$, $p_{2,10}$, $p_{2,11}\}$ and $M\upharpoonright P_{a_2} = \emptyset$).

\begin{definition}[RGER]
Given a KPN $\Sigma = (P_S$, $P_K$, $T$, $F$, $M_0$, $\mathcal{A}$, $L)$ where $\mathcal{A} = \{a_1$, $a_2$, $\cdots$, $a_m\}$, its RGER $\Delta = (\mathbb{M}$, $\mathbb{F}$, $\sim_{a_1}$, $\sim_{a_2}$, $\cdots$, $\sim_{a_m})$ is defined as follows:
\begin{itemize}
\item[1.] $(\mathbb{M}$, $T$, $\mathbb{F})$ is the reachability graph of Petri net $(P_S\cup P_K$, $T$, $F$, $M_0)$; and
\item[2.] $\forall a\in\mathcal{A}$, $\sim_a \subseteq \mathbb{M}\times \mathbb{M}$ is an equivalence relation w.r.t. agent $a$ such that $\forall M$, $M^\prime\in\mathbb{M}$, $M\sim_a M^\prime$ iff $M\upharpoonright P_a = M^\prime~\upharpoonright~P_a$.
\end{itemize}
\end{definition}

In the definition of RGER, we omit transition names on all directed edges because they are not related to our model checking. In fact, $\mathbb{F}\subseteq \mathbb{M}\times \mathbb{M}$ is the set of all directed edges such that $(M$, $M^\prime)\in\mathbb{F}$ iff $\exists t\in T$ such that $M[t\rangle M^\prime$.

For each agent, an equivalence relation is constructed based on the reachability graph. If an agent owns the same basic knowledges at two markings, then the two markings are indistinguishable from the agent's epistemic perspective, i.e., they are equivalent w.r.t. the agent. We call it as an equivalence relation since it is reflexive, symmetric and transitive. Therefore, those markings that are mutually equivalent w.r.t. an agent form an \emph{equivalence class of knowledge} (\emph{equivalence class} for short), and an equivalence relation divides $\mathbb{M}$ into a group of equivalence classes. An equivalence class $Q$ w.r.t. an equivalence relation $\sim_a$ means that at any marking in $Q$, agent $a$ owns the same basic knowledges, i.e., $\forall M$, $M^\prime \in Q: M\upharpoonright P_a=M^\prime\upharpoonright P_a$.

Fig.~\ref{ABPG} shows the reachability graph of the KPN in Fig.~\ref{ABP}. Equivalence relation $\sim_{a_1}$ is shown in Fig. \ref{ABPG1} where an equivalence class is represented in the same color. For example, $\{M_1$, $M_3$, $M_4$, $M_5$, $M_6$, $M_7\}$ is an equivalence class w.r.t. $\sim_{a_1}$ which means that agent $a_1$ only owns one basic knowledge $p_{1,8}$ at these markings (i.e., Alice only knows that she has got a password). Similarly, Fig. \ref{ABPG2} shows equivalence relation $\sim_{a_2}$.

Based on the notions and notations of set and binary relation, we can define \emph{union} $\sim_a\cup\sim_b$, \emph{intersection} $\sim_a\cap\sim_b$ and \emph{transitive closure} $(\sim_a\cup\sim_b)^+$ of two given equivalence relations $\sim_a$ and $\sim_b$. Consequently, $\sim_a\cap\sim_b$ and $(\sim_a\cup\sim_b)^+$ are still equivalence relations; but $\sim_a\cup\sim_b$ is not necessarily an equivalence relation because it is reflexive and symmetric but not necessarily transitive. For example, given equivalence relation $\sim_{a_1}$ in Fig.~\ref{ABPG1} and equivalence relation $\sim_{a_2}$ in Fig.~\ref{ABPG2}, we have that $M_0~(\sim_{a_1}\cup\sim_{a_2})~M_2$ because of $M_0\sim_{a_1}M_2$; and $M_2~(\sim_{a_1}\cup\sim_{a_2})~M_4$ because of $M_2\sim_{a_2}M_4$. However, $(M_0$, $M_4)\not\in(\sim_{a_1}\cup\sim_{a_2})$ because of $(M_0$, $M_4)\not\in\,\,\sim_{a_1}$ and $(M_0$, $M_4)\not\in\,\,\sim_{a_2}$, but $M_0~(\sim_{a_1}\cup\sim_{a_2})^+~M_4$ because of $M_0\sim_{a_1}M_2$ and $M_2\sim_{a_2}M_4$. Besides, $M_1~(\sim_{a_1}\cap\sim_{a_2})~M_3$ because of $M_1\sim_{a_1}M_3$ and $M_1\sim_{a_2}M_3$. These operations of binary relation are important for our definitions of epistemic operators in CTLK.

\section{Symbolic analysis of KPN using OBDD}
For reachability graph, there is the state explosion problem. We use OBDD to deal with it. This section first recalls OBDD~\cite{BDD,CUDD}, then presents our heuristic method of constructing a static variable order in OBDD and finally illustrates our OBDD-based symbolic approaches of producing and encoding the states and relations of an RGER.

For an OBDD, a key factor is the order of variables in it since the order directly decides the scale of the related diagram (tree) and affects the compressed result. Based on the structure and initial marking of a KPN, we propose a heuristic method of ordering variables in OBDD, and thus can reduce the time and space complexity of producing, encoding and exploring the states of an RGER.

An OBDD can symbolically represent all states of a KPN and thus we can significantly reduce the space of storing those states based on a good variable order. Our model checking algorithms (which will be introduced in Section 6) require a part (but not all) of transition or equivalence relations of states in a verification process. To carry out these tasks, we need to utilize OBDD to compute the predecessors or equivalence classes of a given set of states, which will also be introduce in this section.

\subsection{Ordered Binary Decision Diagrams}
OBDD is recalled in this section. For more details, one may refer to~\cite{BDD}. Here, we only review some of its definitions for~readability, but some examples of OBDD can be seen in Section~8.

A Binary Decision Diagram (BDD) is a rooted, directed, and acyclic graph with two sink nodes labelled by 0 or 1 that represent Boolean functions \textbf{0} and \textbf{1}, respectively. Each non-sink node is labelled with a Boolean variable $\upsilon$ and has two out-edges labelled by 1 (that represents \emph{then}) or 0 (that represents \emph{else}). Each non-sink node represents a Boolean function corresponding to its 1-edge if $\upsilon=1$ or a Boolean function corresponding to its 0-edge if $\upsilon=0$.

An OBDD is a BDD where all variables are totally ordered and each path from source node to a sink node visits these variables in the ascending order. A Reduced OBDD (ROBDD) is an OBDD where each node represents a distinct Boolean function and no variable node has identical 1--edge or 0--edge. ROBDD has some important properties. It provides compact representations of Boolean functions. Besides, there are efficient algorithms for performing all kinds of logical operations on ROBDD. They are all based on such a crucial fact that a ROBDD has a canonical representation of a Boolean function: given a fixed variable order, there is exactly one ROBDD representing it for any Boolean function. Notice that we use the ROBDD technique in this paper, but for readability we still call it as OBDD in the next~content.

The OBDD technique can encode large sets of states with small data structures and enable efficient manipulation of those sets. However, it is known that the size of an OBDD for a Boolean function seriously depends on the chosen variable order~\cite{OBDD} and an improper variable order can still result in the node explosion problem, i.e., the number of nodes in an OBDD grows exponentially with the number of variables. To find an optimal variable order is still an NP-hard problem~\cite{MARCIE}, and thus a policy of dynamically reordering variables is often taken. For example, MCMAS takes such a policy so that it often has a very high time complexity since it frequently looks for a good compromise between continuous variable reordering and efficiency of reducing memory-consuming. In this paper, we uses a static variable order instead of dynamically reordering variables. The order is constructed based on the structure and initial marking of a KPN. Therefore, our method of producing and exploring a huge state space encoded by OBDD can save lots of time. In this paper, we use the OBDD-package in the CUDD library~\cite{CUDD} developed by Fabio Somenzi at Colorado University.

\subsection{A Heuristic Method of Ordering Variables in OBDD}
In this paper, we propose a heuristic method to produce a static variable (i.e., place) order based on the structure and initial marking of a KPN.

In an OBDD, a set of states are encoded by a Boolean function composed of variables $x_{p_1}$, $x_{p_2}$, $\cdots$ and $x_{p_n}$. Firing a transition $t$ will change the assignments of these places that belong to $^\bullet t$ $\cup$ $t^\bullet$ (i.e., $x_p$ becomes $\overline{x_p}$ or $\overline{x_p}$ becomes $x_p$). It will spend much time for OBDD to compute a new Boolean function if the distance between two value-changed variables in a variable order is long. Therefore, we can consider the structural and behavioral characteristics of Petri nets and thus reasonably arrange the order of places.

\begin{algorithm}[tt]
\caption{$Order(P_S\cup P_K)$}
\hspace*{0.02in} {\bf Input:}
    KPN $\Sigma$\\
    \hspace*{0.02in} {\bf Output:}
    A static variable order in OBDD
\label{alg:order}
    \begin{algorithmic}[1]
    \State $S=\emptyset$;
    \State $i=0$;
    \While {$(S\neq P_S\cup P_K)$}
    \For {$($each $p\in (P_S\cup P_K)\setminus S)$}
    \If {$(M_0(p)==1)$}
    \State $y_i=x_p$;
    \State $S=S\cup\{p\}$;
    \State $i=i+1$;
    \State break;
    \EndIf
    \EndFor
    \For {$($each $p\in (P_S\cup P_K)\setminus S)$}
    \If {$(\exists t\in T: p\in t^\bullet\wedge \hspace{+0.1mm}^\bullet\hspace{-0.1mm} t\subseteq S)$}
    \State $y_i=x_p$;
    \State $S=S\cup\{p\}$;
    \State $i=i+1$;
    \EndIf
    \EndFor
    \EndWhile
    \State\Return $y_1<y_2<\cdots<y_{|P_S\cup P_K|}$;
    \end{algorithmic}
\end{algorithm}

Two places are \emph{dependent} if there is a transition which affects them~\cite{Order}. Then we can conclude that the shorter the average distance among all dependent places in a variable order, the better effect of compacting the state space. An MAS is a modular system where an agent corresponds to a module and different agents interact/collaborate at some points. In a KPN modelling an MAS, a subnet corresponds to one agent, and different subnets are combined via some common places and thus are loosely coupled. Therefore, we can propose a heuristic method to construct a place (i.e., variable) order. It is described in Algorithm~\ref{alg:order}. First, let $y_1<y_2<\cdots<y_{|P_S\cup P_K|}$, $S$ be the set of places already assigned to some variables and $S=\emptyset$ initially. Second, we randomly choose an unassigned variable $y_i$ and a marked place $p$ at initial marking ($p\not\in S$) and let $y_i=x_p$. Third, we find those places (say $p_j$, $p_k$, $\cdots$) that are not in $S$ but each of them is marked at some marking reached from $S$ ($S$ is seen as a marking) by firing some transition. Let $y_{i+1}=x_{p_j}$, $y_{i+2}=x_{p_k}$, $\cdots$, and we add these places to $S$. We repeat the third step until no place is marked. Then we return the second step and repeat them until $S=P_S\cup P_K$, which means that all places have been assigned. Since $|P_S\cup P_K|$ is limited, Algorithm~1 can be terminated in $\mathcal{O}(|P_S\cup P_K|^2)$.

The order outputted by our algorithm can guarantee that those dependent places in the same subnet (agent) are as close as possible. Due to the feature of loose coupling of different subnets, our algorithm enables that the average distance among all dependent places in this order is short enough so that it can guarantee a good effect of compacting the state space. Later, our experiments will substantiate this~idea.

\subsection{Producing and Encoding All States of a KPN Based on OBDD}
We now present our symbolic approach of producing and encoding all states of a KPN.

Given a KPN, since it is safe, we can use $x_p$ or $\overline{x_p}$ to represent a place $p$ and then a marking $M$ is encoded by logical operation AND (i.e., symbol $\cdot$ in Algorithm~\ref{alg:M}) of places $p_1$, $p_2$, $\cdots$, $p_n$. The assignment of these places is defined as: $p = x_p$ if $M(p)=1$, and $p = \overline{x_p}$ if $M(p)=0$. For example, if there is a KPN where $P=\{p_1$, $p_2$, $p_3$, $p_4\}$, then marking $M=\{p_1$, $p_3\}$ can be represented by $x_{p_1}\cdot\overline{x_{p_2}}\cdot x_{p_3}\cdot\overline{x_{p_4}}$ (or $x_{p_1}\,\overline{x_{p_2}}\, x_{p_3}\,\overline{x_{p_4}}$ for short), which is true only if $x_{p_1}=x_{p_3}=1\wedge x_{p_2}=x_{p_4}=0$. Similarly, a set of markings can be encoded by logical operation OR (i.e., symbol $+$ in Algorithm~\ref{alg:M}) of the corresponding Boolean functions. For example, if $M_1=x_{p_1}\,\overline{x_{p_2}}\, x_{p_3}\,\overline{x_{p_4}}$ and $M_2=x_{p_1}\,\overline{x_{p_2}}\,\overline{x_{p_3}}\,\overline{x_{p_4}}$, then $\{M_1$, $M_2\}=M_1+M_2=x_{p_1}\,\overline{x_{p_2}}\, x_{p_3}\,\overline{x_{p_4}}+x_{p_1}\,\overline{x_{p_2}}\,\overline{x_{p_3}}\,\overline{x_{p_4}}=x_{p_1}\,\overline{x_{p_2}}\,\overline{x_{p_4}}$. It is true only if $x_{p_1}=1\wedge x_{p_2}=x_{p_4}=0$, but the value of $x_{p_3}$ is arbitrary, i.e., $1$ or $0$. It is worthy to note that when a variable does not occur in a Boolean expression, this expression represents such all markings that the place corresponding to the variable is marked or unmarked. These notations and operations refer to the work in~\cite{sy-BPN}.

Based on the advantage that an OBDD can efficiently manipulate sets, we can consider the firing of transitions at a set of markings rather than using the traditional one-by-one marking-producing method. Given a subset $\mathbb{M}_x\subseteq\mathbb{M}$ and a transition $t\in T$, we define two functions: $$Enable(t, \mathbb{M}_x)= \{M\in\mathbb{M}_x\mid M[t\rangle\}$$ which is the markings in $\mathbb{M}_x$ that can enable transition $t$; $$Img(t, \mathbb{M}_x)= \{M\in\mathbb{M}\mid\exists M^\prime\in\mathbb{M}_x: M^\prime[t\rangle M\}$$ which is the markings reachable from $\mathbb{M}_x$ by firing $t$.

Based on the rules of enabling transitions, we can easily calculate $Enable(t$, $\mathbb{M}_x)$, i.e., $$Enable(t, \mathbb{M}_x)=\mathbb{M}_x\cdot\prod\limits_{p\in ^\bullet t}x_p\,.$$ Then we can easily calculate $Img(t$, $\mathbb{M}_x)$ based on the rules of firing transitions. If $Enable(t$, $\mathbb{M}_x)=\emptyset$, then $Img(t$, $\mathbb{M}_x)=\emptyset$. Otherwise, we modify the assignment of places in $Enable(t$, $\mathbb{M}_x)$, i.e., $p = \overline{x_p}$ for each $p\in$ $^\bullet t\setminus t^\bullet$, $p = x_p$ for each $p\in t^\bullet\setminus\hspace{+0.1mm}^\bullet \hspace{-0.1mm}t$, and $p$ is unchanged for other cases. Finally, the modified $Enable(t$, $\mathbb{M}_x)$ is $Img(t$, $\mathbb{M}_x)$.

Based on the function $Img$, we design Algorithm~\ref{alg:M} to produce all reachable markings of a KPN encoded by an OBDD. First of all, $M_0$ is represented by the above encoding rule and the reached markings $Reached=\{M_0\}$ initially. Second, the new markings $New$ are produced by using function $Img$ for each transition $t\in T$ and then $Reached$ is updated by $Reached\,\cup\,New$. We repeat the second step until no new marking is produced. Finally, $Reached$ represents all reachable markings encoded by an OBDD.

\subsection{Computing Predecessors and Equivalence Classes of A Given Set of States}
Based on the output of Algorithm~\ref{alg:M}, we define functions $Pre(\mathbb{M}$, $\mathbb{M}_x)$ to compute the predecessors of markings in $\mathbb{M}_x$ and $Eq(\mathbb{M}$, $\mathbb{M}_x$, $a)$ to compute those equivalent markings to at least one marking in $\mathbb{M}_x$ w.r.t. agent $a$, i.e., the equivalence classes of markings in $\mathbb{M}_x$. They are described in Algorithms~\ref{alg:pre} and~\ref{alg:eq}, respectively.

\begin{algorithm}[tt]
	\caption{$Mark(\Sigma)$}
    \hspace*{0.02in} {\bf Input:}
    KPN $\Sigma$\\
    \hspace*{0.02in} {\bf Output:}
    All markings $\mathbb{M}$ encoded by OBDD
	\label{alg:M}
	\begin{algorithmic}[1]
        \State $M_0 = \textbf{true}$;
        \For {$($each $p\in P_S\cup P_K)$}
        \If {$(M_0(p)==1)$}
        \State $M_0 = M_0\cdot x_p$;
        \Else\,\,$M_0 = M_0\cdot\overline{x_p}$;\EndIf\EndFor
        \State $Reached = From = M_0$;
        \Repeat
        \For {$($each $t\in T)$}
        \State $From = From+Img(t$, $From)$;
        \EndFor
        \State $New = From\setminus Reached$;
        \State $From = New$;
        \State $Reached = Reached+New$;
        \Until {$(New==0)$};
        \State\Return $Reached$;
	\end{algorithmic}
\end{algorithm}

\begin{algorithm}[tt]
\caption{$Pre(\mathbb{M}$, $\mathbb{M}_x)$}
\hspace*{0.02in} {\bf Input:}
    All markings $\mathbb{M}$ and a set of markings $\mathbb{M}_x$\\
    \hspace*{0.02in} {\bf Output:}
    $\{M\in\mathbb{M}\mid \exists t\in T\wedge\exists M^\prime\in\mathbb{M}_x: M[t\rangle M^\prime\}$
\label{alg:pre}
    \begin{algorithmic}[1]
    \State $\mathbb{M}_1=\emptyset$;
    \For {$($each $t\in T)$}
    \State $\mathbb{M}_2=\mathbb{M}_x\cdot\prod\limits_{p\in t^\bullet} x_p $;
    \If {$(\mathbb{M}_2\neq\emptyset)$}
    \State $\mathbb{M}_2$ is updated such that $\forall p\in P_S\cup P_K$\\
    $$ p=\left\{
    \begin{aligned}
    &\overline{x_p}&if\,p\in t^\bullet\setminus\hspace{+0.1mm}^\bullet\hspace{-0.1mm}t\\
    &x_p&if\,p\in\hspace{+0.1mm}^\bullet\hspace{-0.1mm}t\setminus t^\bullet\\
    \end{aligned}\,;
    \right.
    $$
    \State $\mathbb{M}_1=\mathbb{M}_1+\mathbb{M}_2$;
    \EndIf
    \EndFor
    \State\Return $\mathbb{M}\cdot\mathbb{M}_1$;
    \end{algorithmic}
\end{algorithm}

\begin{algorithm}[tt]
\caption{$Eq(\mathbb{M}$, $\mathbb{M}_x$, $a)$}
\hspace*{0.02in} {\bf Input:}
    All markings $\mathbb{M}$, a set of markings $\mathbb{M}_x$ and an agent~$a$\\
    \hspace*{0.02in} {\bf Output:}
    $\{M\in\mathbb{M}\mid \exists M^\prime\in\mathbb{M}_x: M\sim_a M^\prime\}$
\label{alg:eq}
    \begin{algorithmic}[1]
    \State $\mathbb{M}_1=\mathbb{M}_2=\mathbb{M}_x$;
    \For {$($each $p\in(P_S\cup P_K)\setminus P_a)$}
    \State $\mathbb{M}_1$ is updated by $p=x_p$;
    \State $\mathbb{M}_2$ is updated by $p=\overline{x_p}$;
    \State $\mathbb{M}_3=\mathbb{M}_1+\mathbb{M}_2$;
    \State $\mathbb{M}_1=\mathbb{M}_2=\mathbb{M}_3$;
    \EndFor
    \State\Return $\mathbb{M}\cdot\mathbb{M}_3$;
    \end{algorithmic}
\end{algorithm}

\begin{figure} [tt]
\centering
\includegraphics[scale=0.85]{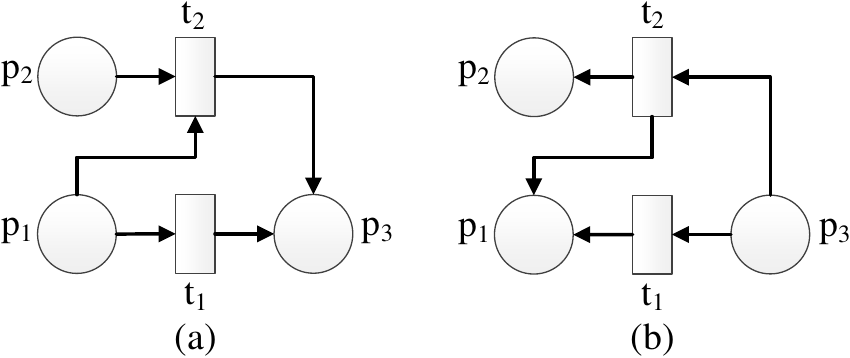}
\caption {(a) A simple net: when it is initialized by marking $\{p_1\}$, it has two reachable markings, i.e., $\{p_1\}$ and $\{p_3\}$; (b) the inversed net of (a): when it is initialized by marking $\{p_3\}$, it has three reachable markings, i.e., $\{p_3\}$, $\{p_1\}$ and $\{p_1$, $p_2\}$.}
\label{PN}
\end{figure}

In fact, computing the predecessors of $\mathbb{M}_x$ in a net is to compute the successors of $\mathbb{M}_x$ in the \emph{inversed net} of the original net. Here the inversed net of a net $(P$, $T$, $F)$ is $(P$, $T$, $F^{-1})$, i.e., the direction of each arc of the original net is inversed. But the following facts should be noted. First, Algorithm~\ref{alg:pre} does not use the inversed net of a KPN, and we only utilize this concept to explain this algorithm. In fact, what we use in the algorithm is the input places of input transitions of places occurring in $\mathbb{M}_x$. Second, every predecessor of $\mathbb{M}_x$ is in these markings computed through this method, but some of these computed markings are possibly not reached in the KPN and thus they should be removed. For removing these fake markings, we only need to implement a set intersection operation of all reachable markings and the computed ones, i.e., the OBDD operation $\mathbb{M}\cdot\mathbb{M}_1$ in Algorithm~\ref{alg:pre}. For example, Fig.~\ref{PN} shows a simply net and its inversed net. When Fig.~\ref{PN} (a) is initialized by marking $\{p_1\}$, there are two reachable markings, i.e., $M_0=\{p_1\}=x_{p_1}\,\overline{x_{p_2}}\, \overline{x_{p_3}}$ and $M_1=\{p_3\}=\overline{x_{p_1}}\,\overline{x_{p_2}}\, x_{p_3}$ where $M_1$ is the successor of $M_0$ and $M_0$ is the predecessor of $M_1$. However, if we do not consider Line~10 in Algorithm~\ref{alg:pre}, i.e., the final $\mathbb{M}_1$ is viewed as the computed predecessors, then we have that $Pre(\mathbb{M}$, $\{M_1\})=x_{p_1}\,\overline{x_{p_2}}\, \overline{x_{p_3}}+x_{p_1}\,x_{p_2}\, \overline{x_{p_3}}$, i.e., it corresponds to the successors $\{p_1\}$ and $\{p_1, p_2\}$ of marking $\{p_3\}$ in the inversed net in Fig.~\ref{PN} (b). Obviously, marking $x_{p_1}\,x_{p_2}\,\overline{x_{p_3}}=\{p_1$, $p_2\}$ is not the predecessor of $M_1$ in Fig.~\ref{PN} (a) when $\{p_1\}$ is the initial marking of this net. $\mathbb{M}\cdot\mathbb{M}_1=x_{p_1}\,\overline{x_{p_2}}\, \overline{x_{p_3}}$ is the right result. Algorithm~\ref{alg:pre} only executes $|T|$ loops and each loop is to execute some simple operations of~OBDD.

It is simple to compute the equivalence classes  of $\mathbb{M}_x$ w.r.t. agent $a$ according to the operations of OBDD. Just stated in Section~4.3, if a variable does not occur in a Boolean expression, the expression represents such all markings in which the place corresponding to the variable is marked or unmarked. Therefore, we only need to eliminate from $\mathbb{M}_x$ those variables corresponding to the places that are not in $P_a$, which represents all possible markings that are equivalent to at least one marking in $\mathbb{M}_x$ w.r.t. agent $a$. However, some of these markings computed by this method are possibly not reached in the KPN, and thus they should be removed. Similar to Algorithm~\ref{alg:pre}, for removing these fake markings, we also implement a set intersection operation of all reachable markings and the computed ones, i.e., the OBDD operation $\mathbb{M}\cdot\mathbb{M}_3$ in Algorithm~\ref{alg:eq}. For example, let $M_0=x_{p_1}\,\overline{x_{p_2}}\,x_{p_3}$, $M_1=x_{p_1}\,x_{p_2}\,\overline{x_{p_3}}$,
$M_2=\overline{x_{p_1}}\,\overline{x_{p_2}}\,\overline{x_{p_3}}$, $P_a=\{p_1\}$, $\mathbb{M}=\{M_0$, $M_1$, $M_2\}=x_{p_1}\,\overline{x_{p_2}}\,x_{p_3}+x_{p_1}\,x_{p_2}\,\overline{x_{p_3}}+\overline{x_{p_1}}\,\overline{x_{p_2}}\,\overline{x_{p_3}}$ and $\mathbb{M}_x=\{M_0$, $M_2\}=x_{p_1}\,\overline{x_{p_2}}\,x_{p_3}+\overline{x_{p_1}}\,\overline{x_{p_2}}\,\overline{x_{p_3}}$. Then we compute $Eq(\mathbb{M}$, $\mathbb{M}_x$, $a)$. First, we eliminate variables $x_{p_2}$ and $x_{p_3}$ from $\mathbb{M}_x$, i.e., $\mathbb{M}_3=x_{p_1}+\overline{x_{p_1}}$. Obviously, $\mathbb{M}_3$ represents 8 markings but some markings such as $x_{p_1}\,x_{p_2}\,x_{p_3}$ are not what we want. $\mathbb{M}\cdot\mathbb{M}_3=x_{p_1}\,\overline{x_{p_2}}\,x_{p_3}+x_{p_1}\,x_{p_2}\,\overline{x_{p_3}}+\overline{x_{p_1}}\,\overline{x_{p_2}}\,\overline{x_{p_3}}$ is the right result. Algorithm~\ref{alg:eq} executes at most $|P|$ loops and each loop is to execute some simple operations of OBDD.

Obviously, Algorithms~\ref{alg:order}--\ref{alg:eq} are all based on the structure characteristics of Petri nets and only use the inputs or outputs of the related places or transitions. The combination of the OBDD technique and the structure characteristics of KPN guarantees that our model checking algorithms can obtain good performances.

\section{CTLK}
We use CTLK~\cite{CTLK} as the specification language of complex knowledges. CTLK extends CTL~\cite{CTL2,CTL,MC1} with epistemic operators so as to reason about the knowledge of agents in MAS. In general, when a kind of modelling language is used to model MAS, the syntax of CTLK is based on this language (e.g., ISPL and KPN) and the induced model (e.g., Kripke model of ISPL and RGER of KPN) is used to explain the semantics of CTLK. Because we use KPN to model MAS, the syntax of CTLK is based on KPN and the semantics of CTLK is based on RGER.

\begin{definition}[Syntax of CTLK]
Given a KPN $\Sigma = (P_S$, $P_K$, $T$, $F$, $M_0$, $\mathcal{A}$, $L)$, the syntax of CTLK is defined by the following Existential Normal Form (ENF) expressions:
\begin{equation}
\begin{split}
\phi ::=\,&\textbf{true} \mid p \mid \neg \phi \mid \phi \wedge \phi \mid EX\,\phi \mid EG\,\phi\mid\\& E\,(\phi\,U\phi)\mid\mathcal{K}_a\mid \mathcal{E}_\Gamma\,\phi\mid \mathcal{D}_\Gamma\,\phi\mid \mathcal{C}_\Gamma\,\phi\nonumber
\end{split}
\end{equation}
where $p\in P_S\cup P_K$, $a\in\mathcal{A}$ and $\Gamma\subseteq \mathcal{A}$.
\end{definition}

Other basic modalities derived from the above ones are defined as follows:
\begin{itemize}
\item[--] $deadlock \stackrel {def}{=}\neg EX\,\textbf{true}$;
\item[--] $\phi_1 \vee \phi_2 \stackrel {def}{=} \neg (\neg \phi_1 \wedge \neg \phi_2)$;
\item[--] $\phi_1 \rightarrow \phi_2 \stackrel {def}{=} \neg \phi_1 \vee \phi_2$;
\item[--] $AX\,\phi \stackrel {def}{=} \neg EX\,\neg \phi\wedge\neg deadlock$;
\item[--] $A\,(\phi_1\, U \,\phi_2)\stackrel {def}{=} \neg E\,(\neg \phi_1\, U \,(\neg \phi_1\wedge \neg \phi_2))\wedge\neg EG(\neg\phi_1)$
\item[--] $EF\,\phi \stackrel {def}{=} E\,(\textbf{true}\, U\, \phi)$;
\item[--] $AG\,\phi \stackrel {def}{=} \neg EF\,\neg \phi$;
\item[--] $AF\,\phi \stackrel {def}{=} A\,(\textbf{true}\,U\, \phi)$.
\end{itemize}

Given an RGER $\Delta = (\mathbb{M}$, $\mathbb{F}$, $\sim_{a_1}$, $\sim_{a_2}$, $\cdots$, $\sim_{a_m})$ and a marking $M\in\mathbb{M}$, a \emph{computation} of $\Delta$ starting from $M$ is a maximal sequence of markings, i.e., $\omega = (M^0$, $M^1$, $\cdots)$ such that $M^0 = M$ and $\forall i\in \mathbb{N}$: $(M^i$, $M^{i+1})\in \mathbb{F}$. A computation may be finite or infinite. For a finite computation $\omega = (M^0$, $M^1$, $\cdots$, $M^n)$, $\omega (i) = M^i$ for each $i\leq n$, and $\omega (i) = \emptyset$ for each $i> n$. For an infinite computation $\omega = (M^0$, $M^1$, $\cdots)$, $\omega (i) = M^i$ for each $i\in\mathbb{N}$. $\Omega(M)$ denotes the set of all computations starting from~$M$.

Since a KPN is safe, we can use one place to represent one atomic proposition in CTLK. A marked place means the value \textbf{true} of the corresponding atomic proposition and otherwise the value \textbf{false}.

\begin{definition} [Semantics of CTLK]
Given an RGER $\Delta = (\mathbb{M}$, $\mathbb{F}$, $\sim_{a_1}$, $\sim_{a_2}$, $\cdots$, $\sim_{a_m})$, a marking $M\in\mathbb{M}$ and a CTLK formula $\phi$, $(\Delta$, $M)\models\phi$ denotes that $\phi$ is true at the marking $M$ in $\Delta$. $\Delta$ can be omitted when no ambiguity takes place. The relation $\models$ is defined inductively as follows:
\begin{itemize}
\item[--]$M\models \textbf{true};$
\item[--]$M\models p$ iff $M(p) = 1;$
\item[--]$M\models \neg \phi$ iff $M \nvDash \phi;$
\item[--]$M\models \phi_1 \wedge \phi_2$ iff $M\models \phi_1$ and $M\models \phi_2;$
\item[--]$M\models EX\,\phi$ iff there exists $\omega\in\Omega(M)$ such that $\omega(1)~\neq~\emptyset\wedge\omega(1)\models\phi;$
\item[--]$M\models EG\,\phi$ iff there exists $\omega\in\Omega(M)$ such that for each $i\in\mathbb{N}$, if $\omega(i)\neq\emptyset$, then $\omega(i)\models\phi;$
\item[--]$M\models E\,(\phi_1\,U\,\phi_2)$ iff there exists $\omega\in\Omega(M)$ and $n\in\mathbb{N}$ such that $\omega(n)\neq\emptyset\wedge\omega(n)\models\phi_2$ and $\omega (j)\models\phi_1$ for each $j\in\{0$, $1$, $\cdots$, $n-1\};$
\item[--]$M\models\mathcal{K}_a\,\phi$ iff $\forall M^\prime\in\mathbb{M}: M^\prime\sim_a M\Rightarrow M^\prime\models\phi;$
\item[--]$M\models\mathcal{E}_\Gamma\,\phi$ iff $\forall M^\prime\in\mathbb{M}: M^\prime~\bigg(\bigcup\limits_{a\in\Gamma}\sim_a\bigg)~M\Rightarrow M^\prime~\models~\phi;$
\item[--]$M\models\mathcal{D}_\Gamma\,\phi$ iff $\forall M^\prime\in\mathbb{M}: M^\prime~\bigg(\bigcap\limits_{a\in\Gamma}\sim_a\bigg)~M\Rightarrow M^\prime~\models~\phi;$
\item[--]$M\models\mathcal{C}_\Gamma\,\phi$ iff $\forall M^\prime\in\mathbb{M}: M^\prime~\bigg(\bigcup\limits_{a\in\Gamma}\sim_a\bigg)^+~M\Rightarrow M^\prime~\models~\phi$.
\end{itemize}
\end{definition}

\begin{definition}[Validity]
A CTLK formula $\phi$ is valid in KPN $\Sigma$ (denoted $\Sigma\models\phi$) if the RGER $\Delta$ of $\Sigma$ satisfies $(\Delta$, $M_0) \models \phi$, i.e., $\phi$ is true at the initial marking of~$\Sigma$.
\end{definition}

As mentioned above, CTLK is an extension of CTL since epistemic operators are considered in it. From Def.~4 we can see that the semantics of the first seven propositions/operators are the same with those of CTL~\cite{CTL2,CTL,MC1}. Here, we only explain the last four epistemic operators.

The semantics of $\mathcal{K}_a$ means that agent $a$ gains the (basic or complex) knowledge $\phi$ (i.e., he knows that $\phi$ is true) at marking $M$ if and only if he can derive that $\phi$ is true at each marking that has the same knowledge with $M$ for agent $a$. In other words, knowledge $\phi$ is true at the equivalence class of marking $M$ w.r.t. agent $a$.

The semantics of $\mathcal{E}_\Gamma$ means that every agent in a set of agents $\Gamma$ gains the knowledge $\phi$ at marking $M$, i.e., $\forall a\in\Gamma: M\models \mathcal{K}_a\,\phi$. Obviously, $\mathcal{K}_a\phi\stackrel{def}{=} \mathcal{E}_{\{a\}}\phi$.

The semantics of $\mathcal{D}_\Gamma$ means that $\phi$ is a distributed knowledge in a set of agents $\Gamma$ at marking $M$, i.e., $M\models\mathcal{K}_{a_\Gamma}\,\phi$ where $a_\Gamma$ is viewed as a special agent who owns all basic knowledges of $\Gamma$. In other words, the basic knowledges of each agents in $\Gamma$ need be collected together to gain knowledge $\phi$. For one agent in $\Gamma$, knowledge $\phi$ is unknowable.

The semantics of $\mathcal{C}_\Gamma$ means that $\phi$ is a common knowledge in a set of agents $\Gamma$ at marking $M$, i.e., $\forall a_{i_1}$, $a_{i_2}$, $a_{i_3}$, $\cdots\in\Gamma: M\models\mathcal{K}_{a_{i_1}}\,\phi\wedge\mathcal{K}_{a_{i_1}}\,\big(\mathcal{K}_{a_{i_2}}\,\phi\big)\wedge\mathcal{K}_{a_{i_1}}\,\big(\mathcal{K}_{a_{i_2}}\, \big(\mathcal{K}_{a_{i_3}}\,\phi\big)\big)\wedge\cdots$. In other words, knowledge $\phi$ can be arbitrarily transitive among $\Gamma$.

For a KPN, each agent can gain some basic knowledges when its basic knowledge places are marked. Therefore, for each $a\in\mathcal{A}$ and each $p\in P_a$, $M\models\mathcal{K}_a\, p$ holds iff $M(p)~=~1$; and similarly, for each $p\in P_K$, $M\models \mathcal{E}_{L(p)}\,p$ holds iff $M(p)=1$.

For an MAS, can an agent (or a set of agents) derive some complex knowledges based on his (or their) basic knowledges? For example in Fig.~\ref{ABP}, after the protocol is executed, do Alice and Bob both know that each one has got the password? CTLK formula $AG\,((p_{1,7}\wedge p_{2,7})\rightarrow\mathcal{E}_{\{a_1, a_2\}}\,(p_{1,8}\wedge p_{2,11}))$ specifies the complex knowledge. Furthermore, is it a common knowledge for Alice and Bob that each other has got the password? CTLK formula $AG\,((p_{1,7}\wedge p_{2,7})\rightarrow \mathcal{C}_{\{a_1, a_2\}}\,(p_{1,8}\wedge p_{2,11}))$ specifies this case. Our model checking method can decide that the first formula is valid but the second one is not, i.e., after the protocol is executed, it is not a common knowledge for Alice and Bob that each other has got the password even though they both know it.

\section{Model Checking Algorithms and Model Checker of CTLK}
In this section, we introduce our model checking algorithms and then show our model checker KPNer.

\begin{algorithm}[tt]
    \caption{$Sat(\Sigma$, $\mathbb{M}$, $\phi)$}
	\hspace*{0.02in} {\bf Input:}
    KPN $\Sigma$, all markings $\mathbb{M}$ and CTLK formula $\phi$\\
	\hspace*{0.02in} {\bf Output:}
    $\{M\in\mathbb{M}\mid M\models\phi\}$
	\label{alg:sat}
	\begin{algorithmic}[1]
        \If {($\phi$ is \textbf{true})} \State\Return $\mathbb{M}$;\EndIf
		\If {($\phi$ is a place)}
        \State\Return $\{M\in \mathbb{M}\mid M(\phi)=1\}$;\EndIf
		\If {($\phi$ is $\neg\phi_1$)} \State\Return $\{M\in \mathbb{M}\mid M\not\in Sat(\Sigma$,~$\mathbb{M}$,~$\phi_1)\}$;\EndIf
        \If {($\phi$ is $\phi_1\wedge\phi_2$)}
        \State\Return $Sat(\Sigma$, $\mathbb{M}$, $\phi_1)\cap Sat(\Sigma$, $\mathbb{M}$, $\phi_2)$;\EndIf
        \If {($\phi$ is $EX\,\phi_1$)} \State\Return $Sat_{EX}(\phi_1)$;\EndIf
        \If {($\phi$ is $EG\,\phi_1$)} \State\Return $Sat_{EG}(\phi_1)$;\EndIf
        \If {($\phi$ is $E$($\phi_1\,U\,\phi_2$))}
        \State\Return $Sat_{EU}(\phi_1$, $\phi_2)$;\EndIf
        \If {($\phi$ is $\mathcal{K}_a\,\phi_1$)}
        \State\Return $Sat_\mathcal{K} (\phi_1$, $a)$;\EndIf
        \If {($\phi$ is $\mathcal{E}_\Gamma\,\phi_1$)} \State\Return $Sat_\mathcal{E} (\phi_1$, $\Gamma)$;\EndIf
        \If {($\phi$ is $\mathcal{D}_\Gamma\,\phi_1$)} \State\Return $Sat_\mathcal{D} (\phi_1$, $\Gamma)$;\EndIf
        \If {($\phi$ is $\mathcal{C}_\Gamma\,\phi_1$)} \State\Return $Sat_\mathcal{C} (\phi_1$, $\Gamma)$;\EndIf
	\end{algorithmic}
\end{algorithm}
\begin{algorithm}[tt]
\caption{$Sat_{EX}\,(\phi)$}
\label{alg:ex}
    \begin{algorithmic}[1]
    \State $X=Sat(\Sigma$, $\mathbb{M}$, $\phi)$;
    \State\Return $Pre(\mathbb{M}$, $X)$;
    \end{algorithmic}
\end{algorithm}
\begin{algorithm}[!ht]
\caption{$Sat_{EG}\,(\phi)$}
\label{alg:eg}
    \begin{algorithmic}[1]
    \State $X_1=\mathbb{M}$; \State $Y_1=Z=Sat(\Sigma$, $\mathbb{M}$, $\phi)$;
    \While {$(X_1\neq Y_1)$}
    \State $X_1=Y_1$;
    \State $Y_1=Z\cap Pre(\mathbb{M}$, $Y_1)$;
    \EndWhile
    \State $X_2=\emptyset$;
    \State $Y_2=Z\cap(\mathbb{M}\setminus Pre(\mathbb{M}$, $\mathbb{M}))$;
    \While {$(X_2\neq Y_2)$}
    \State $X_2=Y_2$;
    \State $Y_2=Y_2\cup(Z\cap Pre(\mathbb{M}$, $Y_2))$;
    \EndWhile
    \State\Return $X_1\cup X_2$;
    \end{algorithmic}
\end{algorithm}
\begin{algorithm}[!ht]
\caption{$Sat_{EU}(\phi_1$,$\phi_2)$}
\label{alg:eu}
    \begin{algorithmic}[1]
    \State $X=\emptyset$;
    \State $Y=Sat(\Sigma$, $\mathbb{M}$, $\phi_2)$;
    \State $Z=Sat(\Sigma$, $\mathbb{M}$, $\phi_1)$;
    \While {$(X\neq Y)$}
    \State $X=Y$;
    \State $Y=Y\cup(Z\cap Pre(\mathbb{M}$, $Y))$;
    \EndWhile
    \State\Return $X$;
    \end{algorithmic}
\end{algorithm}
\begin{algorithm}[!ht]
\caption{$Sat_\mathcal{K} (\phi$, $a)$}
\label{alg:k}
    \begin{algorithmic}[1]
    \State $X=Sat(\Sigma$, $\mathbb{M}$, $\neg\phi)$;
    \State\Return $\mathbb{M}\setminus Eq(\mathbb{M}$, $X$, $a)$;
    \end{algorithmic}
\end{algorithm}
\begin{algorithm}[tt]
\caption{$Sat_\mathcal{E} (\phi$, $\Gamma)$}
\label{alg:e}
    \begin{algorithmic}[1]
    \State $X=Sat(\Sigma$, $\mathbb{M}$, $\neg\phi)$;
    \State $Y=\emptyset$;
    \For {$($each $a\in\Gamma)$}
    \State $Y=Y\cup Eq(\mathbb{M}$, $X$, $a)$;
    \EndFor
    \State\Return $\mathbb{M}\setminus Y$;
    \end{algorithmic}
\end{algorithm}
\begin{algorithm}[!ht]
\caption{$Sat_\mathcal{D} (\phi$, $\Gamma)$}
\label{alg:d}
    \begin{algorithmic}[1]
    \State $X=Sat(\Sigma$, $\mathbb{M}$, $\neg\phi)$;
    \State $Y=\mathbb{M}$;
    \For {$($each $a\in\Gamma)$}
    \State $Y=Y\cap Eq(\mathbb{M}$, $X$, $a)$;
    \EndFor
    \State\Return $\mathbb{M}\setminus Y$;
    \end{algorithmic}
\end{algorithm}
\begin{algorithm}[!ht]
\caption{$Sat_\mathcal{C} (\phi$, $\Gamma)$}
\label{alg:c}
    \begin{algorithmic}[1]
    \State $X=\mathbb{M}$;
    \State $Y=Sat(\Sigma$, $\mathbb{M}$, $\neg\phi)$;
    \While {$(X\neq Y)$}
    \State $X=Y$;
    \For {$($each $a\in\Gamma)$}
    \State $Y=Y\cup Eq(\mathbb{M}$, $Y$, $a)$;
    \EndFor
    \EndWhile
    \State\Return $\mathbb{M}\setminus Y$;
    \end{algorithmic}
\end{algorithm}
\begin{algorithm}[!ht]
    \caption{Model checking algorithm of CTLK}
	\hspace*{0.02in} {\bf Input:}
    KPN $\Sigma$ and CTLK formula $\phi$\\
	\hspace*{0.02in} {\bf Output:}
    $\Sigma\models\phi$ is true or false
	\label{alg:mc}
	\begin{algorithmic}[1]
        \State  $\mathbb{M}=Mark(\Sigma)$;
        \If {$(M_0\in Sat(\Sigma$, $\mathbb{M}$, $\phi))$}
        \State\Return true;
        \EndIf
        \State\Return false;
	\end{algorithmic}
\end{algorithm}

\subsection{Model Checking Algorithms of CTLK}
Since every CTLK formula can be translated into its ENF expression~\cite{MC2}, we just need to give the algorithms of verifying those formulas in ENF.

Our algorithms extend those of CTL in~\cite{MC1}. Given a KPN $\Sigma$ and a CTLK formula $\phi$, the basic verifying procedure mainly includes three steps:
\begin{itemize}
\item[1.] All markings $\mathbb{M}$ in the RGER (i.e., $Mark(\Sigma)$) are produced by Algorithm~\ref{alg:M};
\item[2.] The markings satisfying $\phi$ (i.e., $Sat(\Sigma$, $\mathbb{M}$, $\phi)$) can be computed recursively based on those transition and equivalence relations in the RGER related to $\phi$ that can be computed by Algorithms~\ref{alg:pre} and \ref{alg:eq};
\item[3.] It follows that $\Sigma\models\phi$ if $M_0\in Sat(\Sigma$, $\mathbb{M}$, $\phi)$.
\end{itemize}

Algorithm~\ref{alg:sat} describes a high-level structure of recursively computing $Sat(\Sigma$, $\mathbb{M}$, $\phi)$.

The algorithms $Sat_{EX}$, $Sat_{EG}$ and $Sat_{EU}$ of temporal operators $EX$, $EG$ and $EU$ are described in Algorithms~\ref{alg:ex}, \ref{alg:eg} and \ref{alg:eu}, respectively. The algorithms $Sat_\mathcal{K}$, $Sat_\mathcal{E}$, $Sat_\mathcal{D}$ and $Sat_\mathcal{C}$ of epistemic operators $\mathcal{K}$, $\mathcal{E}$, $\mathcal{D}$ and $\mathcal{C}$ are described in Algorithms~\ref{alg:k}, \ref{alg:e}, \ref{alg:d} and \ref{alg:c}, respectively.

Algorithm~\ref{alg:ex} shows the process of computing $Sat_{EX}\,(\phi)$. First, we compute $Sat(\Sigma$, $\mathbb{M}$, $\phi)$. Second, we use function $Pre$ to compute all predecessors of $Sat(\Sigma$, $\mathbb{M}$, $\phi)$ and these predecessors is $Sat_{EX}\,(\phi)$.

Algorithm~\ref{alg:eg} shows the process of computing $Sat_{EG}\,(\phi)$. There are two cases: the computations without deadlock and the computations with deadlock. For the former, we first compute $Sat(\Sigma$, $\mathbb{M}$, $\phi)$ and then use function $Pre$ to iteratively compute $Sat(\Sigma$, $\mathbb{M}$, $\phi\wedge EX\,\phi)$, $Sat(\Sigma$, $\mathbb{M}$, $\phi\wedge EX\,(\phi\wedge EX\,\phi))$, $\cdots$ until the result no longer changes. Then the final result satisfies $EG\,\phi$. For the latter, we use function $Pre$ to iteratively compute $Sat(\Sigma$, $\mathbb{M}$, $\phi\wedge \neg(EX\,\textbf{true}))$, $Sat(\Sigma$, $\mathbb{M}$, $\phi\wedge EX(\phi\wedge \neg(EX\,\textbf{true})))$, $\cdots$ until the result no longer changes. Notice that $\neg(EX\,\textbf{true})\equiv deadlock$. Then the union of these sets satisfies $EG\,\phi$. Finally, the union of the two cases is $Sat_{EG}\,(\phi)$.

\begin{figure*} [!ht]
\centering
\includegraphics[width=\textwidth]{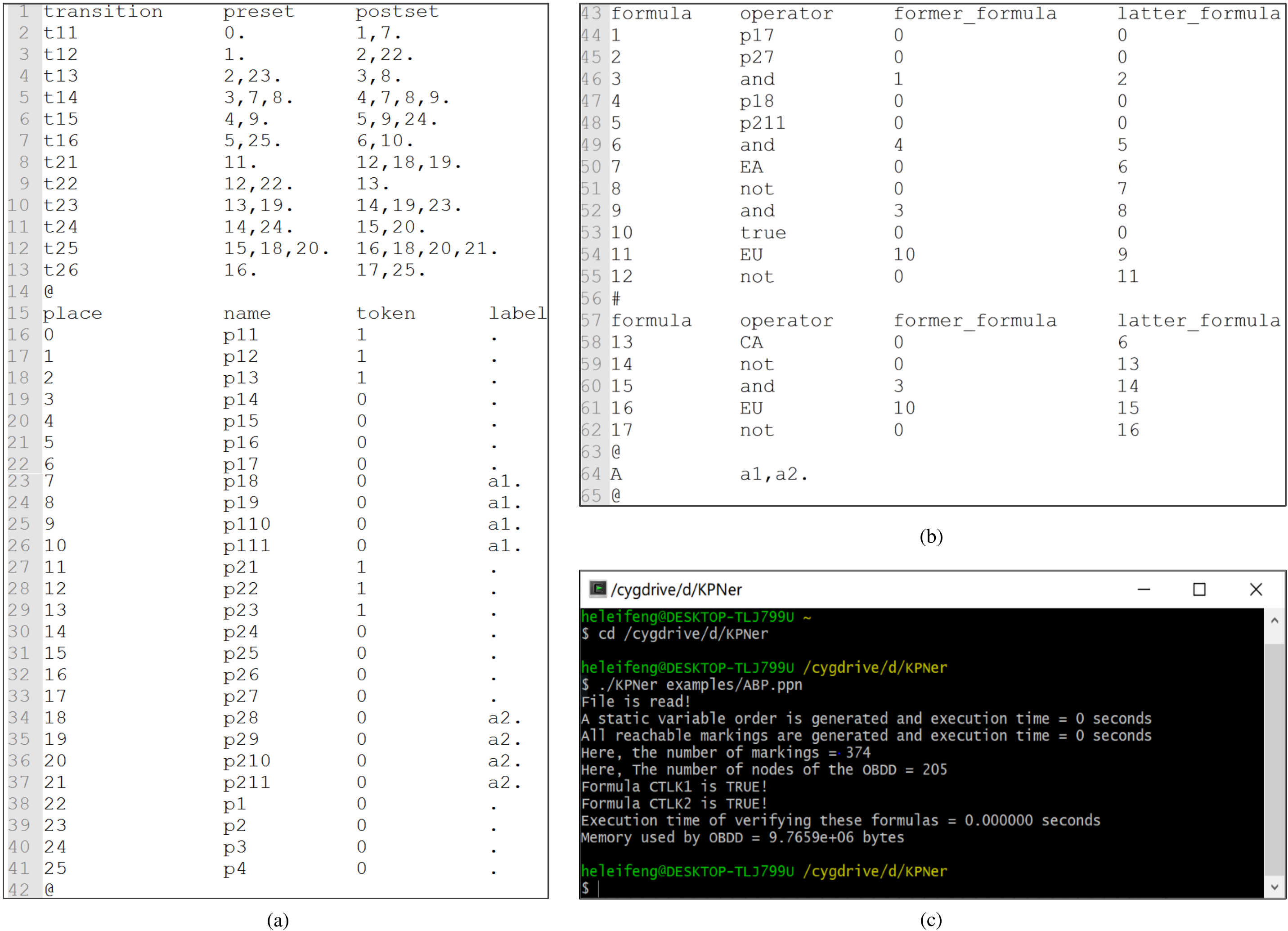}
\caption {(a) The specification of the KPN in Fig.~\ref{ABP}; (b) the specifications of formulas $\phi_1$ and $\phi_2$; (c) the verification results.}
\label{ppn}
\end{figure*}

Algorithm~\ref{alg:eu} shows the process of computing $Sat_{EU}(\phi_1$, $\phi_2)$. First, we compute $Sat(\Sigma$, $\mathbb{M}$, $\phi_1)$ and $Sat(\Sigma$, $\mathbb{M}$, $\phi_2)$. Second, we use function $Pre$ to iteratively compute $Sat(\Sigma$, $\mathbb{M}$, $\phi_1\wedge EX\,\phi_2)$, $Sat(\Sigma$, $\mathbb{M}$, $\ \phi_1\wedge EX\,(\phi_1\wedge EX\,\phi_2))$, $\cdots$ until the result no longer changes. Then the union of $Sat(\Sigma$, $\mathbb{M}$, $\phi_2)$ and these sets is $Sat_{EU}(\phi_1$,$\phi_2)$.

Algorithm~\ref{alg:k} shows the process of computing $Sat_\mathcal{K} (\phi$, $a)$. Because it is not easy to directly compute this set, we choose to compute its complement set, i.e., $\neg Sat_\mathcal{K} (\phi$, $a)$. First, we compute $Sat(\Sigma$, $\mathbb{M}$, $\neg\phi)$. Second, we use function $Eq$ to look for those markings that are equivalent to at least one marking in $Sat(\Sigma$, $\mathbb{M}$, $\neg\phi)$ w.r.t. agent $a$, and all those markings are exactly $\neg Sat_\mathcal{K} (\phi$, $a)$. Then the complement of this set is $Sat_\mathcal{K} (\phi$, $a)$.

Algorithm~\ref{alg:e} shows the process of computing $Sat_\mathcal{E}(\phi$, $\Gamma)$. Similarly, we compute its complement set, i.e., $\neg Sat_\mathcal{E}(\phi$, $\Gamma)$. First, we compute $Sat(\Sigma$, $\mathbb{M}$, $\neg\phi)$. Second, we use function $Eq$ to look for those markings that are equivalent to at least one marking in $Sat(\Sigma$, $\mathbb{M}$, $\neg\phi)$ w.r.t. at least one agent in $\Gamma$, and all these markings are exactly $\neg Sat_\mathcal{E} (\phi$, $\Gamma)$. Then the complement of this set is $Sat_\mathcal{E} (\phi$, $\Gamma)$.

Algorithm~\ref{alg:d} shows the process of computing $Sat_\mathcal{D} (\phi$, $\Gamma)$. Similarly, we compute its complement set, i.e., $\neg Sat_\mathcal{D} (\phi$, $\Gamma)$. First, we compute $Sat(\Sigma$, $\mathbb{M}$, $\neg\phi)$. Second, we use function $Eq$ to look for those markings that are equivalent to at least one marking in $Sat(\Sigma$, $\mathbb{M}$, $\neg\phi)$ w.r.t. each agent in $\Gamma$, and all these markings are exactly $\neg Sat_\mathcal{D} (\phi$, $\Gamma)$. Then the complement of this set is $Sat_\mathcal{D} (\phi$,~$\Gamma)$.

Algorithm~\ref{alg:c} shows the process of computing $Sat_\mathcal{C} (\phi$, $\Gamma)$. Similarly, we compute its complement set, i.e., $\neg Sat_\mathcal{C} (\phi$, $\Gamma)$. First, we compute $Sat(\Sigma$, $\mathbb{M}$, $\neg\phi)$. Second, we use function $Eq$ to iteratively look for those markings that can access at least one marking in $Sat(\Sigma$, $\mathbb{M}$, $\neg\phi)$ via a finite sequence of equivalence relations w.r.t. those agents in $\Gamma$, and all these markings are exactly $\neg Sat_\mathcal{C} (\phi$, $\Gamma)$. Then the complement of this set is $Sat_\mathcal{C} (\phi$, $\Gamma)$.

Algorithm~\ref{alg:mc} describes our model checking process.

For example in Fig. \ref{ABP}, we can use our algorithms to verify the following formulas:

$\phi_1=AG\,((p_{1,7}\wedge p_{2,7})\rightarrow \mathcal{E}_{\{a_1, a_2\}}\,(p_{1,8}\wedge p_{2,11}))$;

$\phi_2=AG\,((p_{1,7}\wedge p_{2,7})\rightarrow \mathcal{C}_{\{a_1, a_2\}}\,(p_{1,8}\wedge p_{2,11}))$.

We first verify $\phi_1$. Its ENF expression is
$$\neg E\,(\textbf{true}\,U (p_{1,7}\wedge p_{2,7}\wedge \neg \mathcal{E}_{\{a_1,a_2\}}\, (p_{1,8}\wedge p_{2,11}))).$$

Based on the RGER in Figs. \ref{ABPG1} and \ref{ABPG2} (all states have been produced but the related transition and equivalence relations of states are computed when they are required), our recursive algorithms obtain the following sets in turns:
\begin{itemize}
\item[1.] Compute $Sat(\Sigma$, $\mathbb{M}$, $\mathcal{E}_{\{a_1,a_2\}}\, (p_{1,8}\wedge p_{2,11}))$:
\begin{itemize}
\item[1.1.] $Sat(\Sigma$, $\mathbb{M}$, $\neg (p_{1,8}\wedge p_{2,11}))=\{M_1$, $M_2$, $\cdots$, $M_{10}$, $M_{11}\}$;
\item[1.2.] $Sat(\Sigma$, $\mathbb{M}$, $\neg\mathcal{E}_{\{a_1,a_2\}}\, (p_{1,8}\wedge p_{2,11}))=\{M_1$, $M_2$, $\cdots$, $M_{12}$, $M_{13}\}$;
\item[1.3.] $Sat(\Sigma$, $\mathbb{M}$, $\mathcal{E}_{\{a_1,a_2\}}\, (p_{1,8}\wedge p_{2,11}))=\{M_{14}\}$;
\end{itemize}
\item[2.] $Sat(\Sigma$, $\mathbb{M}$, $\neg\mathcal{E}_{\{a_1,a_2\}}\, (p_{1,8}\wedge p_{2,11}))=\{M_1$, $M_2$, $\cdots$, $M_{12}$, $M_{13}\}$;
\item[3.] $Sat(\Sigma$, $\mathbb{M}$, $p_{1,7}\wedge p_{2,7})=\{M_{14}\}$;
\item[4.] $Sat(\Sigma$, $\mathbb{M}$, $p_{1,7}\wedge p_{2,7}\wedge\neg\mathcal{E}_{\{a_1,a_2\}}\, (p_{1,8}\wedge p_{2,11}))=\emptyset$;
\item[5.] $Sat(\Sigma$, $\mathbb{M}$, $E\,(\textbf{true}\,U (p_{1,7}\wedge p_{2,7}\wedge\neg\mathcal{E}_{\{a_1,a_2\}}\,(p_{1,8}\wedge p_{2,11}))))=\emptyset$;
\item[6.] $Sat(\Sigma$, $\mathbb{M}$, $\neg E\,(\textbf{true}\,U (p_{1,7}\wedge p_{2,7}\wedge\neg\mathcal{E}_{\{a_1,a_2\}}\, (p_{1,8}\wedge p_{2,11}))))=\mathbb{M}$.
\end{itemize}

Therefore, we have that $M_0\in Sat(\Sigma$, $\mathbb{M}$, $\phi_1)$ and thus $\Sigma\models\phi_1$. Similarly, we have that $\Sigma\not\models\phi_2$.

The complexity of our model checking algorithms consists of two parts. First, all states must be produced, but the number of states possibly grows exponentially even though KPN is safe. Therefore, we use OBDD to encode these states instead of explicitly representing them. The size of an OBDD is only related to the corresponding Boolean function, the number of variables and the variable order, but not related to the number of states. But for some worst cases, OBDD can still have the node explosion problem. Second, we should consider the complexity of verifying CTLK. As shown in~\cite{MC2}, for a CTL formula $\phi$ and a labelled transition system LTS with $n$ states and $k$ transition relations, the CTL model-checking problem LTS $\models\phi$ can by determined in time $\mathcal{O} ((n+k)\cdot |\phi|)$ where $|\phi|$ is the number of atomic propositions and operators in $\phi$. When we also use the uncompressed RGER to verify CTLK in~\cite{KPN}, the complexity of the related algorithm is linear w.r.t. $n+k+w$ and $|\phi|$, i.e., $\mathcal{O}((n+k+w)\cdot |\phi|)$, where $n$ is the number of all states, $k$ is the number of all transition relations of states and $w$ is the number of all equivalence relations of states. Certainly, $n$ grows exponentially at the worst case and $w$ and $k$ grow more seriously than $n$. When we use OBDD to verify CTLK in this paper, the complexity of the related algorithm (i.e., Algorithm~\ref{alg:mc}) depends on the complexity of the operations of OBDD, the size of OBDD and the length of formulas. Certainly, at the worst case the size of OBDD still grows exponentially~\cite{MARCIE} if the variable order is terrible, but lots of studies~\cite{sy-BPN,MCMAS,MARCIE} show that OBDD can work well at most of cases.

\subsection{Model Checker of CTLK}
We develop a model checker KPNer written in C++ programming language based on the above algorithms.

After inputting a KPN and one or more CTLK formulas, KPNer can output the verification results. KPN and formulas are stored in a \emph{.ppn} file, and KPNer can read them. Fig.~\ref{ppn} (a) shows the specification of the KPN in Fig.~\ref{ABP}, (b) shows the specifications of formulas $\phi_1$ and $\phi_2$, and (c) shows the verification results. The results show that $\phi_1$ is valid but $\phi_2$ is invalid, which is coincident with the above analysis. Time spent on verifying them is very short: $<0.000001$ s.

\section{Application}
\subsection{Verification of Alice-Bob Protocol}
\begin{figure} [tt]
\centering
\includegraphics[width=0.485\textwidth]{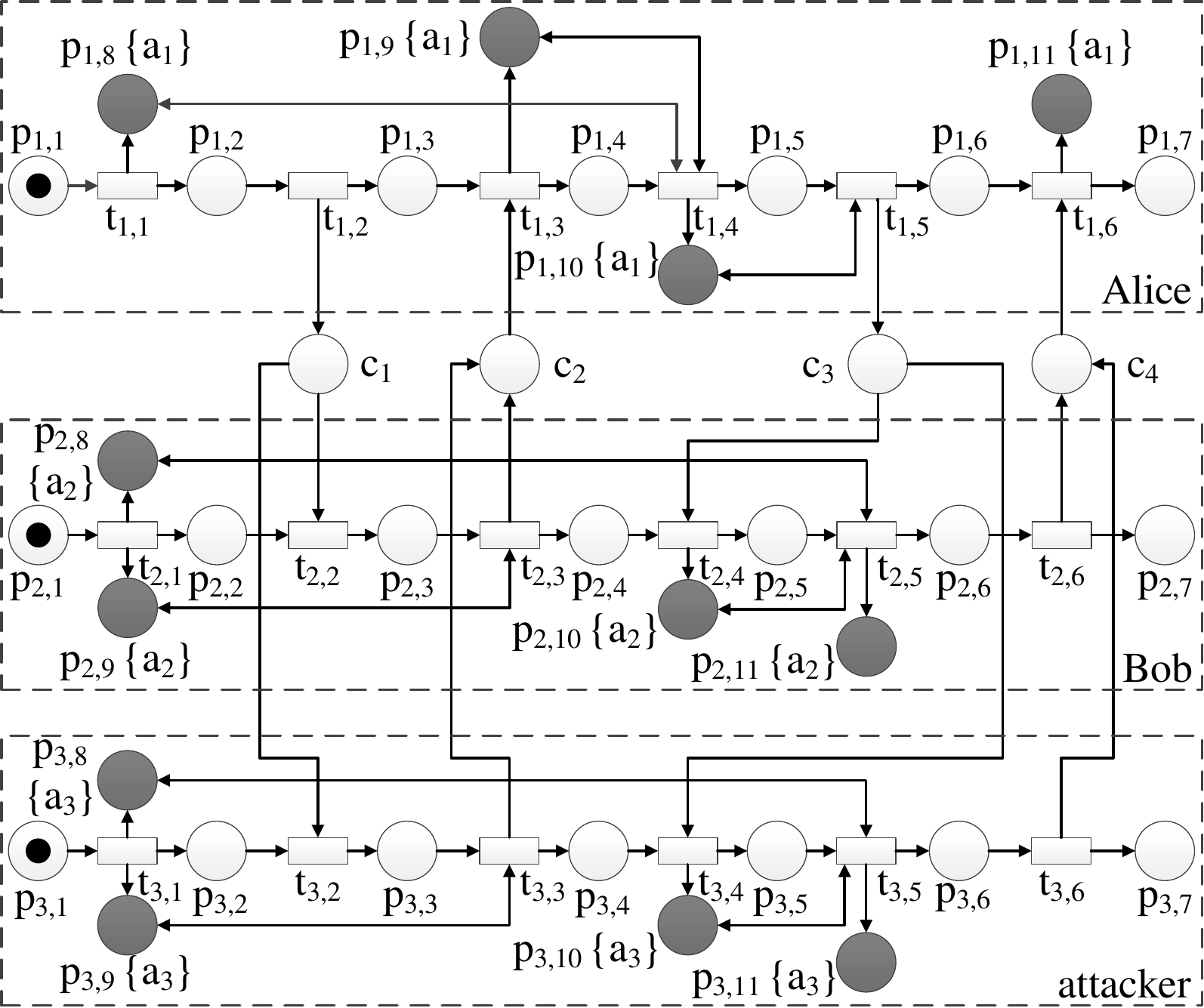}
\caption {KPN model of Alice-Bob Protocol with an attacker.}
\label{ABPT}
\end{figure}
\begin{figure*}[tt]
\centering
\includegraphics[scale=.19]{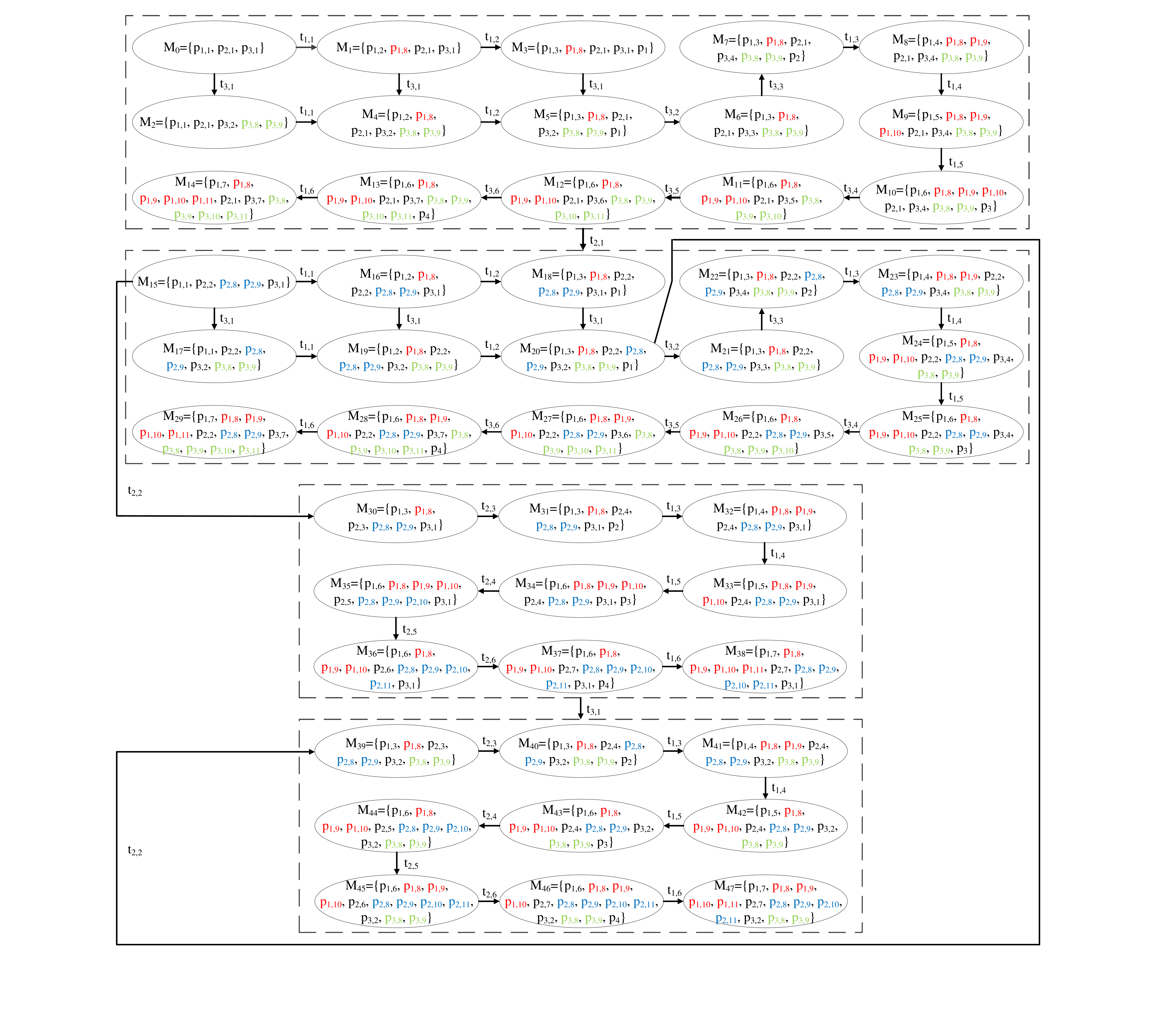}
\caption {The reachability graph of the KPN in Fig.~\ref{ABPT}. Note that the two dotted boxes in the top have an edge labeled by $t_{2,1}$ means that there is an edge from $M_{i}$ to $M_{i+15}$ labeled by $t_{2,1}$ for each $i\in\{0$, $1$, $\cdots$, $14\}$. Similarly, the two dotted boxes in the bottom have an edge labeled by $t_{3,1}$ means that there is an edge from $M_{i}$ to $M_{i+9}$ labeled by $t_{3,1}$ for each $i\in\{30$, $31$, $\cdots$, $38\}$. Figs.~\ref{ABPTG1} and~\ref{ABPTG2} are both similar with this figure.}
\label{ABPTG}
\end{figure*}
\begin{figure*}[tt]
\centering
\includegraphics[scale=.19]{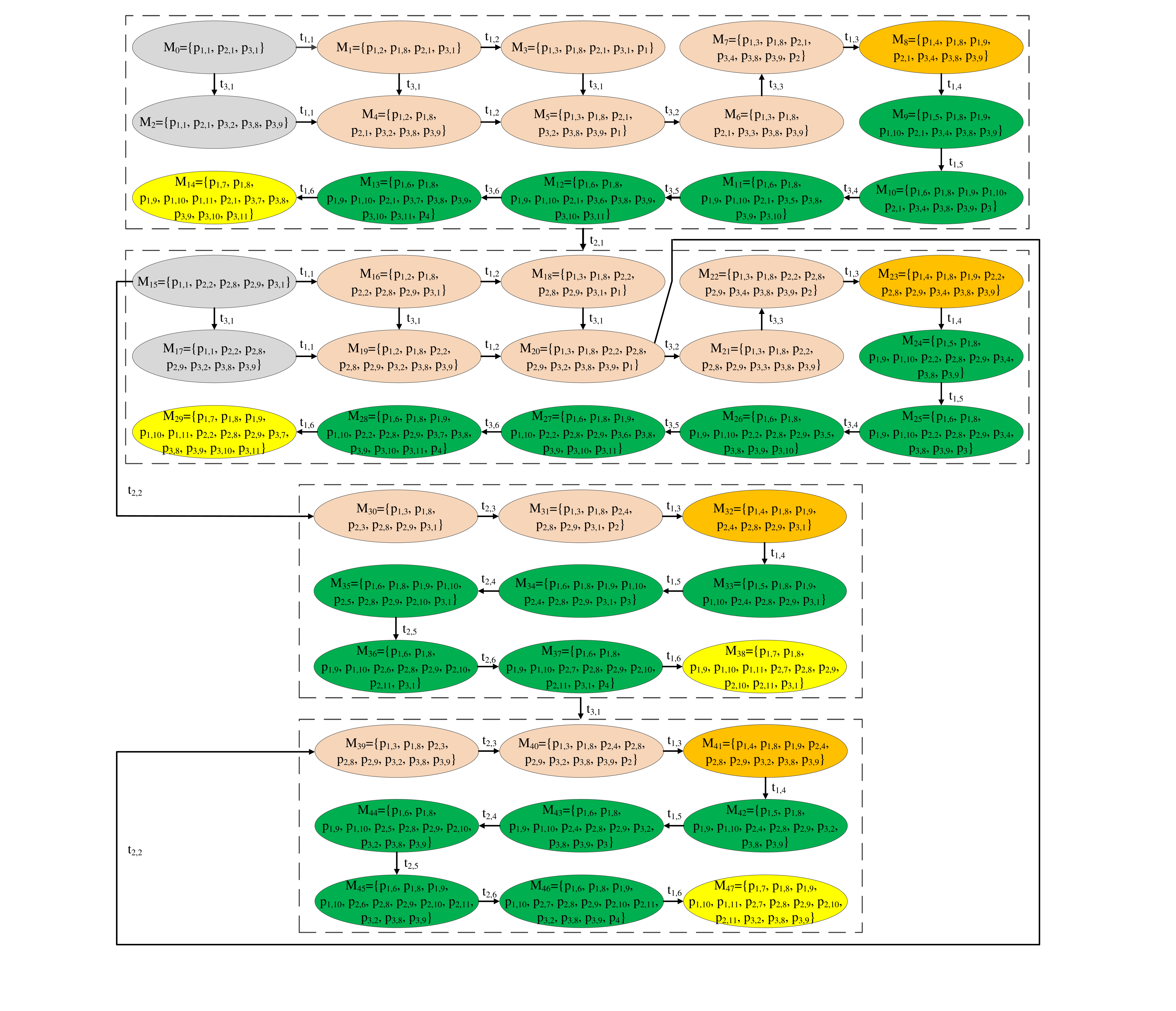}
\caption {Equivalence relation $\sim_{a_1}$ in Fig.~\ref{ABPTG}.}
\label{ABPTG1}
\end{figure*}
\begin{figure*}[tt]
\centering
\includegraphics[scale=.19]{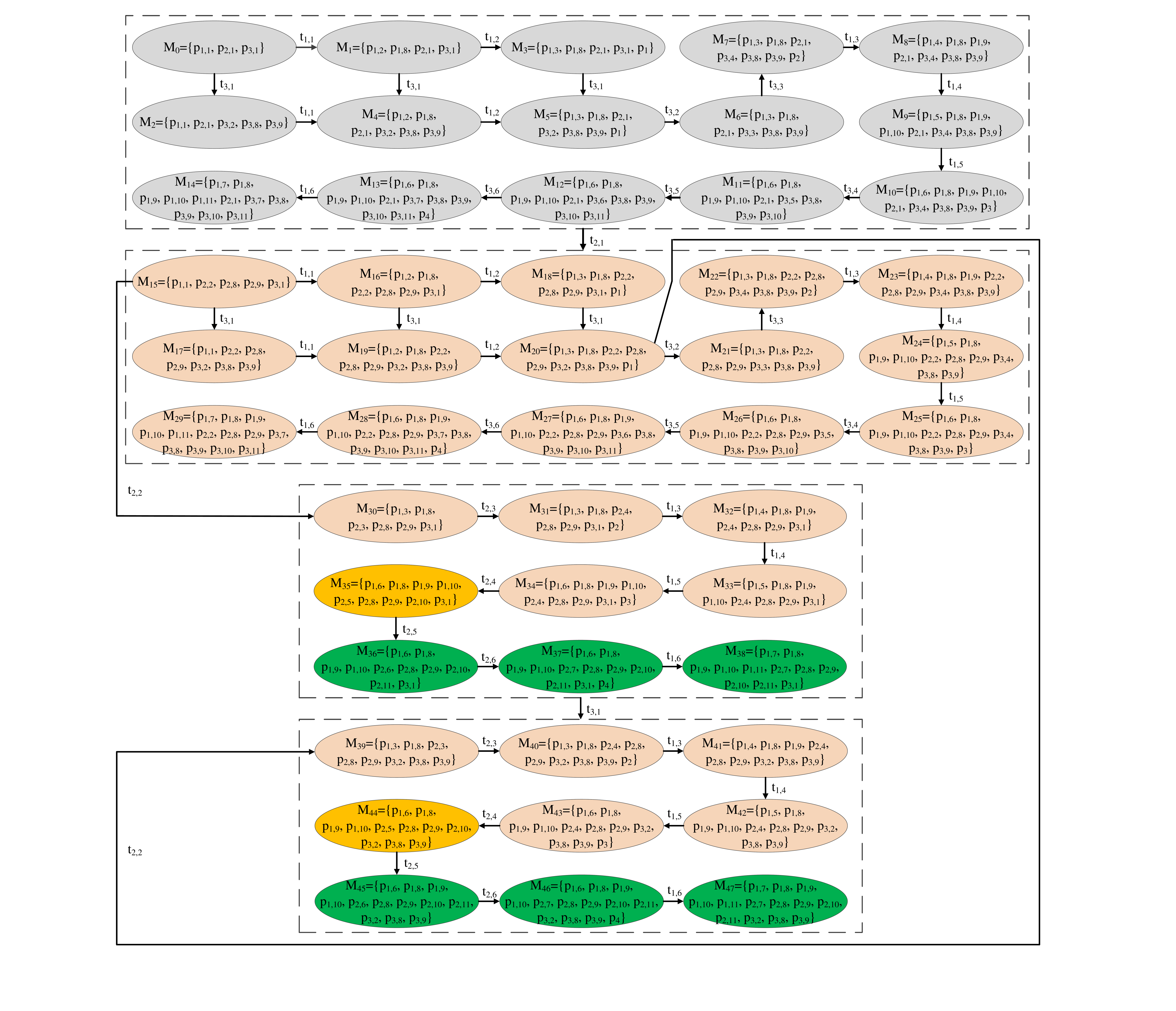}
\caption {Equivalence relation $\sim_{a_2}$ in Fig.~\ref{ABPTG}.}
\label{ABPTG2}
\end{figure*}
\begin{figure*} [tt]
\centering
\includegraphics[width=\textwidth]{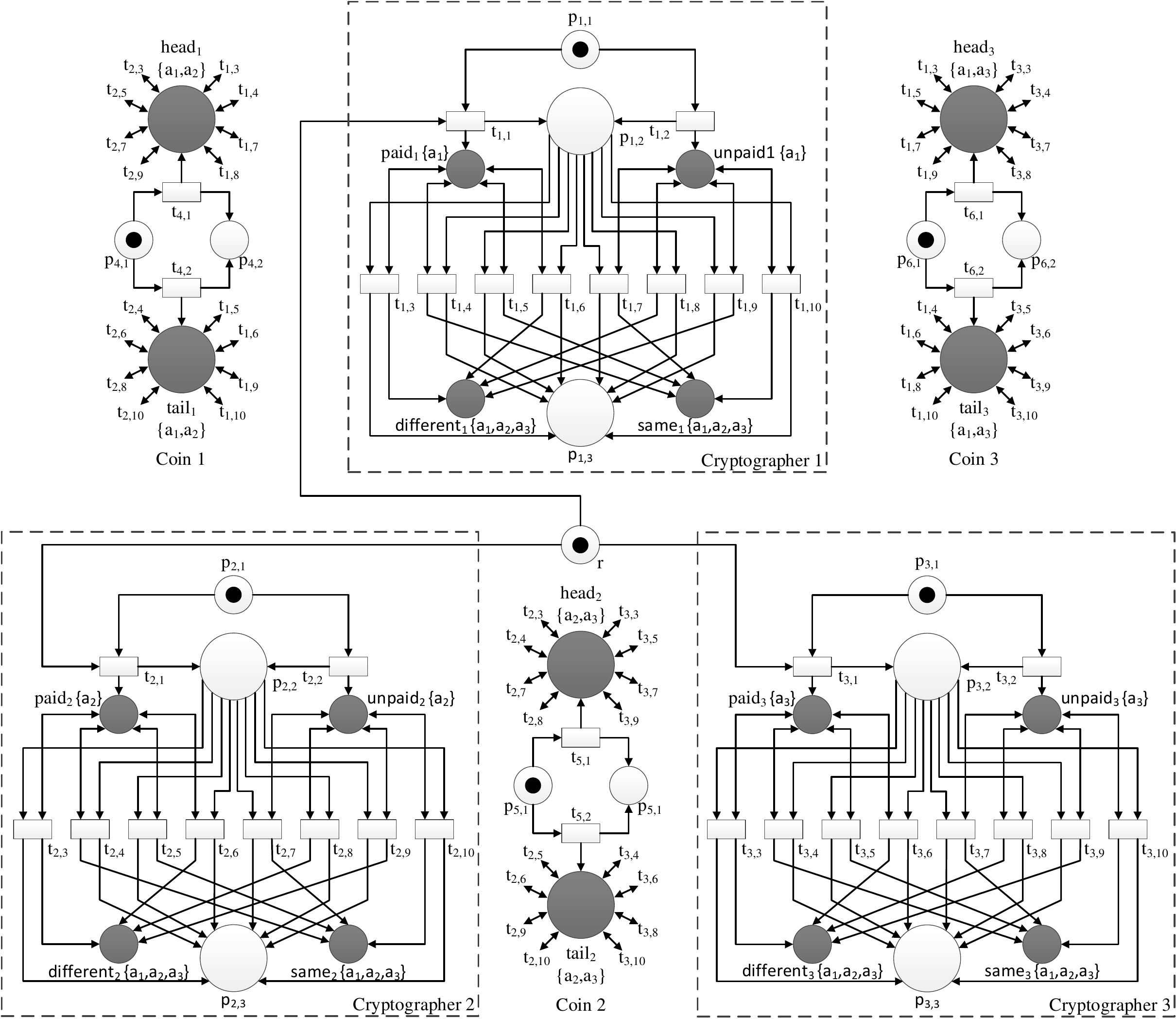}
\caption {KPN model of Dining Cryptographers Protocol in parallel pattern.}
\label{DCP}
\end{figure*}

Alice-Bob Protocol is a famous communication protocol. As shown in~Section~6, it can achieve the basic communication requirement, i.e., one knows that the other has received the password when the protocol is executed. However, it becomes insecure when the messages in the channel are intercepted by an attacker and then the attacker completely copies what Bob does.

Fig.~\ref{ABPT} models this protocol with an attacker. First, the attacker (agent $a_3$) intercepts the request sent from Alice to Bob ($t_{3,2}$) and sends his public key to Alice ($t_{3,3}$). Second, the attacker intercepts the password encrypted by Alice ($t_{3,4}$), uses his private key to decrypt it ($t_{3,5}$) and thus gets the password ($p_{3,11}$). Finally, the attacker sends an acknowledge to Alice ($t_{3,6}$).

In this protocol, the security is destroyed due to the following facts:
\begin{itemize}
\item[1.] The attacker can finally get the password; and
\item[2.] Alice and Bob do not know that the password is decoded by an attacker.
\end{itemize}

They can be specified by the following formula:
$$\phi_3=EF\,p_{3,11}\wedge\neg EF\,(K_{a_1}\,p_{3,11})\wedge\neg EF\,(K_{a_2}\,p_{3,11})$$

Based on our algorithms, we can check $\phi_3$ so as to prove the insecurity of Alice-Bob Protocol. The ENF expression of $\phi_3$ is

\begin{equation}
\begin{split}
&E\,(\textbf{true}\,U\,p_{3,11})\wedge\neg E\,(\textbf{true}\,U\,(K_{a_1}\,p_{3,11}))\,\wedge\\&\neg E\,(\textbf{true}\,U\,(K_{a_2}\,p_{3,11}))\nonumber
\end{split}
\end{equation}

We can produce the RGER of the KPN in Fig.~\ref{ABPT}. The reachability graph is shown in Fig.~\ref{ABPTG}. Because $\phi_3$ is related to the knowledge of agent $a_1$ and $a_2$, we only give equivalence relations $\sim_{a_1}$ and $\sim_{a_2}$ as shown in Figs.~\ref{ABPTG1} and~\ref{ABPTG2}, respectively.

Then our recursive algorithms can compute the following sets in turn:

\begin{itemize}
\item[1.] Compute $Sat(\Sigma$, $\mathbb{M}$, $E\,(\textbf{true}\,U p_{3,11}))$:
\begin{itemize}
\item[1.1.] $Sat(\Sigma$, $\mathbb{M}$, $p_{3,11})=\{M_{12}$, $M_{13}$, $M_{14}$, $M_{27}$, $M_{28}$, $M_{29}\}$;
\item[1.2.] $Sat(\Sigma$, $\mathbb{M}$, $E\,(\textbf{true}\,U p_{3,11}))=\{M_0$, $M_1$, $\cdots$, $M_{28}$, $M_{29}\}$;
\end{itemize}
\item[2.] Compute $Sat(\Sigma$, $\mathbb{M}$, $\neg E\,(\textbf{true}\,U (K_{a_1}\,p_{3,11})))$:
\begin{itemize}
\item[2.1.] $Sat(\Sigma$, $\mathbb{M}$, $\neg p_{3,11})=\{M_0$, $\cdots$, $M_{11}$, $M_{15}$, $\cdots$, $M_{26}$, $M_{30}$, $\cdots$, $M_{47}\}$;
\item[2.2.] $Sat(\Sigma$, $\mathbb{M}$, $\neg K_{a_1}\,p_{3,11})=\mathbb{M}$;
\item[2.3.] $Sat(\Sigma$, $\mathbb{M}$, $K_{a_1}\,p_{3,11})=\emptyset$;
\item[2.4.] $Sat(\Sigma$, $\mathbb{M}$, $E\,(\textbf{true}\,U (K_{a_1}\,p_{3,11})))=\emptyset$;
\item[2.5.] $Sat(\Sigma$, $\mathbb{M}$, $\neg E\,(\textbf{true}\,U (K_{a_1}\,p_{3,11})))=\mathbb{M}$;
\end{itemize}
\item[3.] Compute $Sat(\Sigma$, $\mathbb{M}$, $\neg E\,(\textbf{true}\,U K_{a_2}\,p_{3,11}))$:
\begin{itemize}
\item[3.1.] $Sat(\Sigma$, $\mathbb{M}$, $\neg K_{a_2}\,p_{3,11})=\mathbb{M}$;
\item[3.2.] $Sat(\Sigma$, $\mathbb{M}$, $K_{a_2}\,p_{3,11})=\emptyset$;
\item[3.3.] $Sat(\Sigma$, $\mathbb{M}$, $E\,(\textbf{true}\,U (K_{a_2}\,p_{3,11})))=\emptyset$;
\item[3.4.] $Sat(\Sigma$, $\mathbb{M}$, $\neg E\,(\textbf{true}\,U (K_{a_2}\,p_{3,11})))=\mathbb{M}$;
\end{itemize}
\item[4.] $Sat(\Sigma$, $\mathbb{M}$, $E\,(\textbf{true}\,U p_{3,11})\wedge\neg E\,(\textbf{true}\,U (K_{a_1}\,p_{3,11}))\wedge\neg E\,(\textbf{true}\,U (K_{a_2}\,p_{3,11})))=\{M_0$, $M_1$, $\cdots$, $M_{28}$, $M_{29}\}$.
\end{itemize}

Finally, we can find $M_0\in Sat(\Sigma$, $\mathbb{M}$, $\phi_3)$ and thus $\Sigma\models\phi_3$. Similarly, we can also find $\Sigma\not\models\phi_1$ because Alice can no longer derive that Bob has got the password even though she has received an acknowledgement (this acknowledge may also come from the attacker). This means that Alice-Bob Protocol is not secure while this kind of security is not related with the specific encryption and decryption
technique taken in it. The analysis results outputted by our tool are identical with the above formal calculus and~derivation.

\subsection{Verification of Dining Cryptographers Protocol}
Anonymity protocols are a class of protocols aiming at establishing the privacy of principals during an exchange. One well-known example is \emph{Dining Cryptographers Protocol}~\cite{Din3}. In this protocol, $n$ ($n\geq 3$) cryptographers share a meal around a circular table, and either one of them or their employer pays for the meal. If one of them paid, they would like to discover whether one of them paid without revealing the identity of the payer. Otherwise, they all know their employer paid.

To this end, there is a coin between any two cryptographers. The coin between two cryptographers is randomly tossed and the result (i.e., head or tail) can only be seen by the two cryptographers but cannot be seen by others. This protocol requires
that each cryptographer makes an announcement (i.e., say ``same'' or ``different'' of the two coins beside him or her). If a cryptographer paid money, he or she tells a lie, else he or she tells the truth. After all cryptographer make an announcement, everyone knows whether the employer paid or one of cryptographers paid, but he or she cannot identify the cryptographer who paid unless the payer is himself or herself.

\begin{figure*} [tt]
\centering
\includegraphics[width=\textwidth]{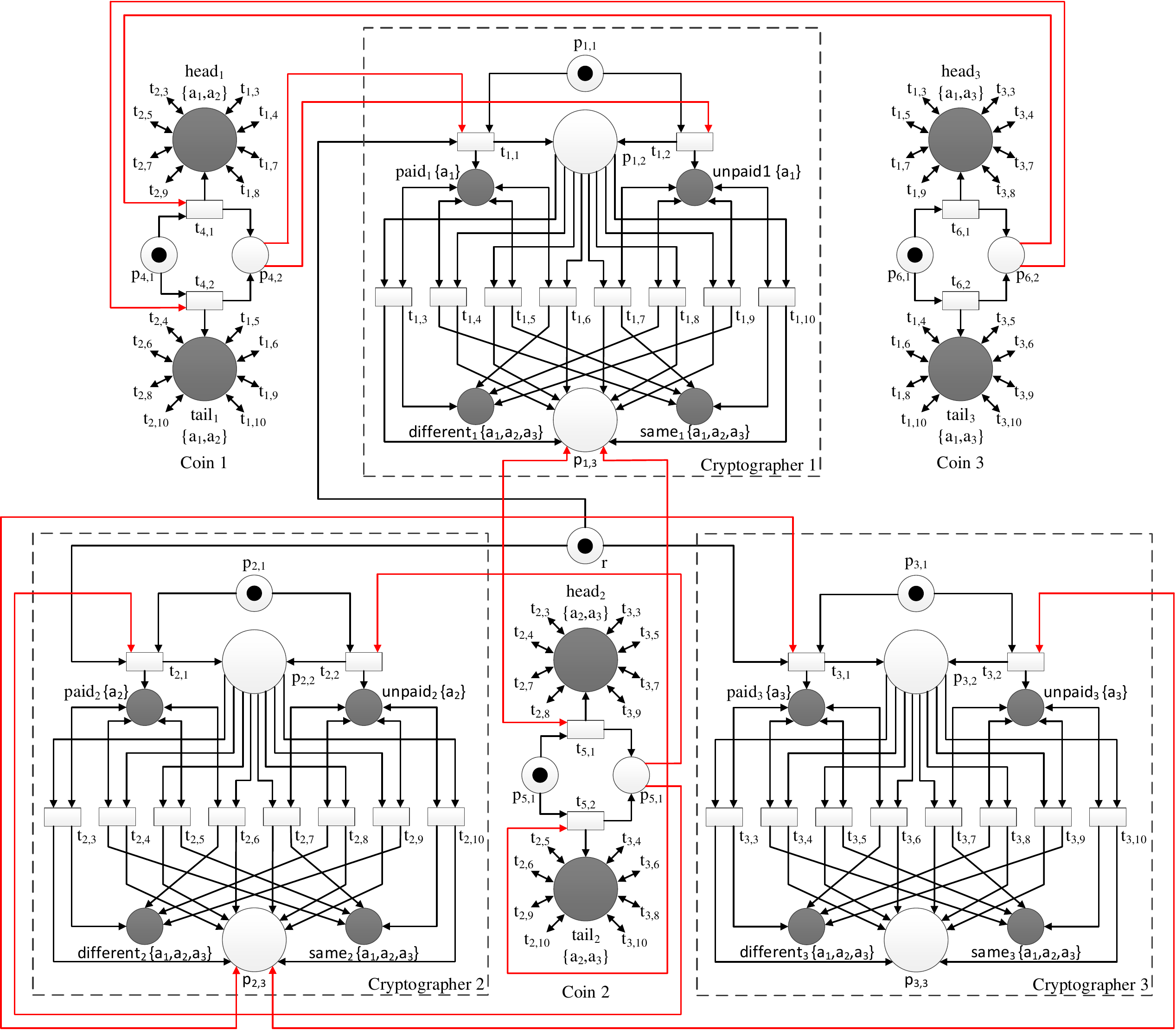}
\caption {KPN model of Dining Cryptographers Protocol in sequential pattern.}
\label{DCP2}
\end{figure*}

Our model and method can prove that this protocol indeed achieves the requirement. The KPN in Fig.~\ref{DCP} models this protocol for the case of 3 cryptographers and is simpler than the KPN in our conference paper~\cite{KPN}. For more details, one may refer to~\cite{KPN}. Here, we does not repeat it. In the next section, we will use this protocol as the benchmark to do the comparison experiments.

\section{Experiments and Comparison}
\renewcommand\arraystretch{1.2}
\begin{table*}[tt]
\footnotesize
\caption{Basic information of the benchmark in parallel and sequential patterns and to-be-checked formulas}
\label{tab-input}
\tabcolsep 9.46pt 
\begin{threeparttable}
\begin{tabular*}{\textwidth}{ccccccccccc}
\toprule
  \multirow{3}*{\makecell{No. of \\cryptos\\$(n)$}}&\multicolumn{3}{c}{KPNs in the parallel pattern}&\multicolumn{3}{c}{KPNs in the sequential pattern}&\multicolumn{2}{c}{$\phi_4$}&\multicolumn{2}{c}{$\phi_5$}\\
\cmidrule(r){2-4} \cmidrule(r){5-7} \cmidrule(r){8-9} \cmidrule(r){10-11}
~&$|P_S\cup P_K|$&$|T|$&$|F|$&$|P_S\cup P_K|$&$|T|$&$|F|$&$|AP|_{\phi_4}$&$|OP|_{\phi_4}$&$|AP|_{\phi_5}$ &$|OP|_{\phi_5}$\\
~&$(11n+1)$&$(12n)$&$(45n)$&$(11n+1)$&$(12n)$&$(49n-2)$&$(3n)$&$(3n+7)$&$(2n+1)$&$(5)$\\\hline
  $10$&$111$&$120$&$450$&$111$&$120$&$488$&$30$&$37$&$21$&$5$\\
  $20$&$221$&$240$&$900$&$221$&$240$&$978$&$60$&$67$&$41$&$5$\\
  $30$&$331$&$360$&$1350$&$331$&$360$&$1468$&$90$&$97$&$61$&$5$\\
  $32$&$353$&$384$&$1440$&$353$&$384$&$1566$&$96$&$103$&$65$&$5$\\
  $34$&$375$&$408$&$1530$&$375$&$408$&$1664$&$102$&$109$&$69$&$5$\\
  $36$&$397$&$432$&$1620$&$397$&$432$&$1762$&$108$&$115$&$73$&$5$\\
  $38$&$419$&$456$&$1710$&$419$&$456$&$1860$&$114$&$121$&$77$&$5$\\
  $40$&$441$&$480$&$1800$&$441$&$480$&$1958$&$120$&$127$&$81$&$5$\\
  $100$&$1101$&$1200$&$4500$&$1101$&$1200$&$4898$&$300$&$307$&$201$&$5$\\
  $200$&$2201$&$2400$&$9000$&$2201$&$2400$&$9798$&$600$&$607$&$401$&$5$\\
  $300$&$3301$&$3600$&$13500$&$3301$&$3600$&$14698$&$900$&$907$&$601$&$5$\\
  $400$&$4401$&$4800$&$18000$&$4401$&$4800$&$19598$&$1200$&$1207$&$801$&$5$\\
  $500$&$5501$&$6000$&$22500$&$5501$&$6000$&$24498$&$1500$&$1507$&$1001$&$5$\\
  $600$&$6601$&$7200$&$27000$&$6601$&$7200$&$29398$&$1800$&$1807$&$1201$&$5$\\
  $700$&$7701$&$8400$&$31500$&$7701$&$8400$&$34298$&$2100$&$2107$&$1401$&$5$\\
  $800$&$8801$&$9600$&$36000$&$8801$&$9600$&$39198$&$2400$&$2407$&$1601$&$5$\\
  $900$&$9901$&$10800$&$40500$&$9901$&$10800$&$44098$&$2700$&$2707$&$1801$&$5$\\
  $1000$&$11001$&$12000$&$45000$&$11001$&$12000$&$48998$&$3000$&$3007$&$2001$&$5$\\
  $1100$&$12101$&$13200$&$49500$&$12101$&$13200$&$53898$&$3300$&$3307$&$2201$&$5$\\
  $1200$&$13201$&$14400$&$54000$&$13201$&$14400$&$58798$&$3600$&$3607$&$2401$&$5$\\
  $1300$&$14301$&$15600$&$58500$&$14301$&$15600$&$63698$&$3900$&$3907$&$2601$&$5$\\
\bottomrule
\end{tabular*}
\begin{itemize}[leftmargin=*]
\item $|AP|_{\phi}$ means the number of atomic propositions in formula $\phi$.
\item $|OP|_{\phi}$ means the number of operators in formula $\phi$.
\end{itemize}
\end{threeparttable}
\end{table*}
\renewcommand\arraystretch{1.2}
\begin{table*}[!ht]
\footnotesize
\caption{Experimental results of KPNer and MCMAS for the benchmark in the sequential pattern.}
\label{tab-K-M}
\tabcolsep 6.69pt 
\begin{threeparttable}
\begin{tabular*}{\textwidth}{cccccccccc}
\toprule
  \multirow{2}*{\makecell{No. of\\ cryptos\\($n$)}}&\multicolumn{5}{c}{KPNer}&\multicolumn{4}{c}{MCMAS}\\
\cmidrule(r){2-6} \cmidrule(r){7-10}
~&$|\mathbb{M}|$&$T_{11}$ (s)&$T_{12}$ (s)&$T_{13}$ (s)&OBDD memory (B)&No. of states&$T_{21}$ (s)&$T_{22}$ (s)&OBDD memory (B)\\\hline
  $10$&$75783$&$<0.001$&$0.031$&$0.032$&$1.392\times 10^7$&$45056$&$2.16$&$0.022$&$1.729\times 10^7$\\
  $20$&$1.615\times 10^8$&$0.016$&$0.187$&$0.36$&$3.383\times 10^7$&$8.808\times 10^7$&$9.454$&$0.466$&$8.83\times 10^7$\\
  $30$&$2.513\times 10^{11}$&$0.031$&$0.562$&$1.219$&$6.048\times 10^7$&--&Timeout&--&--\\
  $32$&$1.074\times 10^{12}$&$0.046$&$0.673$&$1.453$&$5.966\times 10^7$&$5.669\times 10^{11}$&$109.512$&$2.523$&$1.785\times 10^8$\\
  $34$&$4.57\times 10^{12}$&$0.063$&$0.781$&$1.765$&$5.813\times 10^7$&--&Timeout&--&--\\
  $36$&$1.938\times 10^{13}$&$0.078$&$0.906$&$2.125$&$5.668\times 10^7$&$1.017\times 10^{13}$&$15587.3$&$28.192$&$5.584\times 10^8$\\
  $38$&$8.191\times 10^{13}$&$0.078$&$1.062$&$2.5$&$5.633\times 10^7$&--&Timeout&--&--\\
  $40$&$3.452\times 10^{14}$&$0.093$&$1.188$&$3.031$&$5.752\times 10^7$&--&Timeout&--&--\\
  $100$&INF&$1.359$&$17.547$&$113.829$&$7.406\times 10^7$&--&Timeout&--&--\\
\bottomrule
\end{tabular*}
\begin{itemize}[leftmargin=*]
\item $T_{11}$ means the time spent by KPNer to construct a variable order in OBDD; $T_{12}$ means the time spent by KPNer to produce and encode all reachable markings; $T_{13}$ means the time spent by KPNer to verify $\phi_4$ and $\phi_5$; $T_{21}$ means the time spent by MCMAS to produce and encode a complete Kripke model; $T_{22}$ means the time spent by MCMAS to verify $\phi_4$ and $\phi_5$.
\item -- means that the result is not outputted.
\item Timeout means that the time is more than 12 hours.
\item INF means that the number of states in an OBDD cannot be counted up.
\end{itemize}
\end{threeparttable}
\end{table*}

In this section, we use Dining Cryptographers Protocol as the benchmark to do the comparison experiments. First, we compare our tool KPNer with the state-of-the-art CTLK model checker MCMAS. Second, we show the performances of KPNer on two kinds of patterns: parallel pattern of $n$ cryptographers and sequential pattern of $n$ cryptographers. Finally, we introduce a classical heuristic method~\cite{heu} of constructing a static variable order and compare it with our heuristic method.

\subsection{Benchmark}
Dining Cryptographers Protocol was often used as a benchmark~\cite{MCK,MCTK,MCMAS} because it can be expanded to the case of more cryptographers. The following two epistemic requirements were also considered in~\cite{MCK,MCTK,MCMAS}: 1) when each cryptographer has said ``same'' or ``different'', everyone either knows that the employer paid, or knows that one cryptographer paid but cannot know who paid; 2) when each cryptographer has said ``same'' or ``different'' and the employer paid, it is a common knowledge for all cryptographers that the employer paid. Due to symmetry, we only consider the related knowledge to Cryptographer 1 when verifying the first requirement. The two requirements can be formalised by the following formulas:

$\phi_4=AG\bigg(\bigg(\bigwedge\limits_{i=1}^n c^i_{said}\wedge\neg c^1_{paid}\bigg)\rightarrow\bigg(\mathcal{K}_{c_1}\,e_{paid}\vee\bigg(\mathcal{K}_{c_1}\,\bigg(\bigvee\limits_{i=2}^n c^i_{paid}\bigg)\wedge\bigwedge\limits_{i=2}^n \neg\mathcal{K}_{c_1}\,c^i_{paid}\bigg)\bigg)\bigg)$,

$\phi_5=AG\bigg(\bigg(\bigwedge\limits_{i=1}^n c^i_{said}\wedge e_{paid}\bigg)\rightarrow \mathcal{C}_\mathcal{A}\,e_{paid}\bigg)$,

\noindent where $n$ is the number of cryptographers, $c_1$ represents Cryptographer 1, $c^i_{said}$ represents that Cryptographer $i$ said ``same'' or ``different'', $c^i_{paid}$ represents that Cryptographer $i$ paid, and $e_{paid}$ represents that the employer paid.

When modelling this protocol, we use those KPNs whose structures are similar to the KPN in Fig.~\ref{DCP} and obviously these cryptographers are in a parallel pattern. But the related models used in the MCMAS~\cite{MCMAS} are in a sequential pattern. Therefore, for relatively fair comparisons, we also consider a sequential pattern. We just need to add several arcs into the KPN in Fig.~\ref{DCP} to get a sequential pattern, as shown in Fig.~\ref{DCP2}. The idea is as follows: Coin~3 is first tossed, then Coin~1 is tossed, then Cryptographer~1 goes to see the tossed result and say ``same'' or ``different'', then Coin~2 is tossed, then Cryptographer~2 goes to do so, and finally Cryptographer~3 goes to do so. TABLE~\ref{tab-input} shows basic information of these two patterns and formulas for different $n$ (i.e., the number of cryptographers).

\begin{figure*}[ht]
\centering
\includegraphics[width=\textwidth]{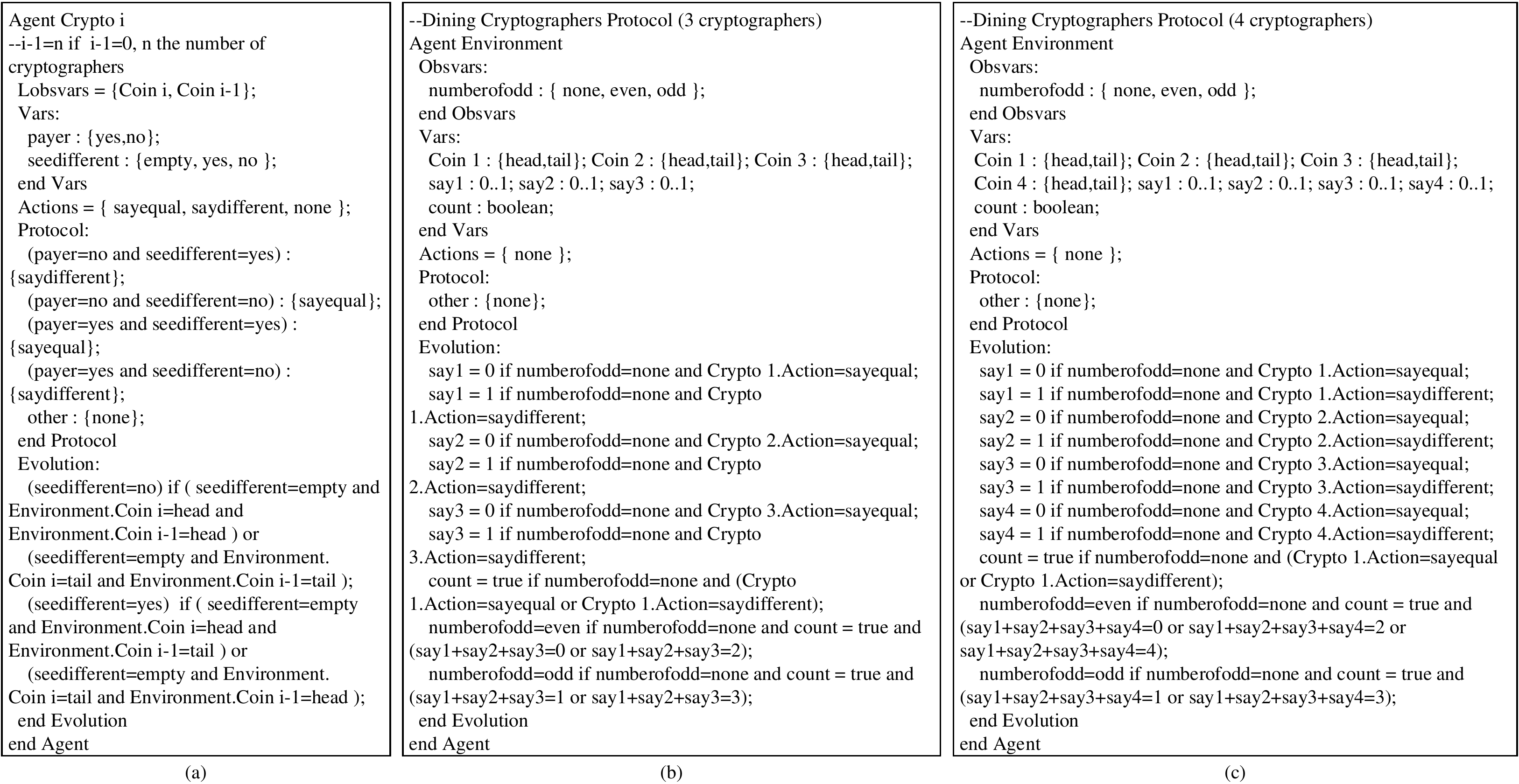}
\caption {A part program of ISPL describing Dining Cryptographers Protocol.}
\label{ISPL2}
\end{figure*}

\subsection{Comparison of KPNer and MCMAS}
To the best of our knowledge, there are three CTLK model checkers MCK~\cite{MCK}, MCTK~\cite{MCTK} and MCMAS~\cite{MCMAS}. They are similar but their difference is that they use different modelling languages to describe MAS. As shown in~\cite{MCMAS}, MCMAS has the best performance among them. Therefore, in this paper we only compare KPNer with MCMAS.

For MCMAS, there is almost no concurrency in ISPL, because it is not allowed to execute a local action of one agent independently and all executable local actions must constitute a joint action so that they can be executed synchronously. Therefore, in order to present a fair comparison, we use the sequential pattern like Fig.~\ref{DCP2} to compare KPNer with MCMAS. Besides, the two model checkers both use the same CUDD version 2.5.1~\cite{CUDD} for a fair comparison. TABLE~\ref{tab-K-M} shows the experimental results. The results show that $\phi_4$ and $\phi_5$ are both valid in this protocol. Here, we only show their performances through increasing the number of cryptographers.

The results show that KPNer is much more efficient and needs less memory than MCMAS. MCMAS can only verify this protocol for the case of up to 36 cryptographers in 12 hours but KPNer can verify this protocol for the case of 100 cryptographers in 3 minutes. Obviously, KPNer is able to handle larger number of cryptographers. Later, we will show that KPNer can verify this protocol: up to 1200 cryptographers in 14 hours for the parallel pattern and up to 600 cryptographers in 14 hours for the sequential pattern. Notice that CUDD cannot count up the number of states in an OBDD when the OBDD is composed of more than 1024 Boolean variables. Therefore, when the number of cryptographers is more than or equal to 100 (the number of places $>1024$), we cannot know the number of markings in $\mathbb{M}$. Here, we use INF (infinite) to represent these numbers. Besides, the performance of MCMAS is unstable. For example, when the number of cryptographers is 30 or 34, MCMAS spends more than 12 hours to produce a complete Kripke model but does not output any result. However, when the number of cryptographers is 32 or 36, MCMAS can output the verification results in 5 hours. The performance of KPNer is stable due to a good static variable order. Because our method to construct a variable order is heuristic, it is also possible for KPNer to find a better order in some larger KPNs. This is the reason why OBDD memory consumed for 32 cryptographers is less than 30 cryptographers' for KPNer. But this improvement is so small that the time still increases with increasing the number of cryptographers. Note that all experiments are conducted on a PC equipped with Inter(R) Core(TM) i5-9400F CPU @ 2.90GHz and RAM @ 16.00G.

We can also see that the performance of KPNer is closely related to the time of verification but the performance of MCMAS is closely related to the time of generating a Kripke model from an ISPL program. This is because KPNer needs to produce the related transition and equivalence relations of states only when verifying a formula. But for MCMAS, all transition and equivalence relations of states are produced before verification, which is time-consuming. When we verify a few formulas, it is unnecessary to do so because verifying a formula only needs to construct the transition and equivalence relations of a part of states in an intermediate model (i.e., Kripke model or RGER). This is one reason why KPNer is much more efficient than MCMAS. For example, KPNer can verify this protocol for the case of 36 cryptographers in 4 seconds but MCMAS spends about 4.3 hours to do so. At the same time, OBDD memory consumed by MCMAS in this process is larger than KPNer's. Besides, we think that the following two reasons also ensure that KPNer outperforms MCMAS:

\begin{itemize}
\item[1.] Each state in Kripke model is global so that it first needs to produce the local state of each agent and then combine these local states into a joint one according to its environment agent. Additionally, for each agent, when producing a new local state from the current local state, the agent considers not only its actions but also the actions of others due to the synchronous semantics required by ISPL. Therefore, it spends much time to produce a new state from a current state. Fortunately, our KPNer only needs to check the pre-set and post-set of a transition when producing a new state from a current state, which can save much time.
\item[2.] MCMAS dynamically reorders variables when OBDD produces and encodes Kripke model. It needs much time to find a good compromise between continuous variable reordering and efficiency of reducing memory-consuming. This is also the reason why the performance of MCMAS is unstable. Fortunately, our KPNer can construct a static variable order based on the structure and initial marking of a KPN and the experiments have shown that these orders are good enough to encode the state space and reduce memory-consuming, so much time can be saved too.
\end{itemize}

Additionally, we use this protocol to show the complexity of the construction of one environment agent in MCMAS. Fig.~\ref{ISPL2} shows a part program of ISPL describing this protocol where (a) describes one cryptographer, (b) describes the environment agent for 3 cryptographers and (c) describes the environment agent for 4 cryptographers. Obviously, when scaling up this protocol, the environment agent uses more complex sentences. This is mainly because it needs to control two kind of common variables, i.e., global variables observable by all agents (e.g., numberofodd) and local variables observable by at least two agents (e.g., Coin i observable by Cryptographer i-1 and Cryptographer i). But our KPNs can easily control these variables by labeling function.

\subsection{Experiments of KPNer for Parallel and Sequential Patterns}
We use KPNer to verify $\phi_4$ and $\phi_5$ for the two patterns like Figs.~\ref{DCP} and~\ref{DCP2}, and TABLE~\ref{tab-K-K} shows the experimental results.

The results show that KPNer performs much better for the parallel pattern than the sequential one. For the parallel pattern, KPNer can verify the case of up to 1200 cryptographers in 14 hours. But for the sequential pattern, it can only verify the case of up to 600 ones in 14 hours. These results are surprising because the state explosion problem of the parallel pattern is much more serious than the sequential pattern's. For example, we estimate that the number of states of 600 cryptographers is about $10^{180}$ in the sequential pattern but is about $10^{540}$ in the parallel pattern. We think that it is because our heuristic method is more suitable for those KPNs with a high degree of concurrency so that the constructed variable order is good enough to offset such serious state explosion problem. For the parallel pattern, different subnets (agent) are loosely coupled and thus randomly finding a few marked places in one subnet can ensure that most of its subsequent places in the same subnet are continuously add to the order from near to far. In other words, it keeps those dependent places in the same subnet as close as possible so that the average distance among all dependent places is short. Therefore, it leads to a very good variable order. However, for the sequential pattern, almost all transitions are limited to be fired in a fixed order and thus randomly finding a few marked places in one subnet cannot usually bring about the above good~result.

\begin{figure} [tt]
\centering
\includegraphics[scale=0.74]{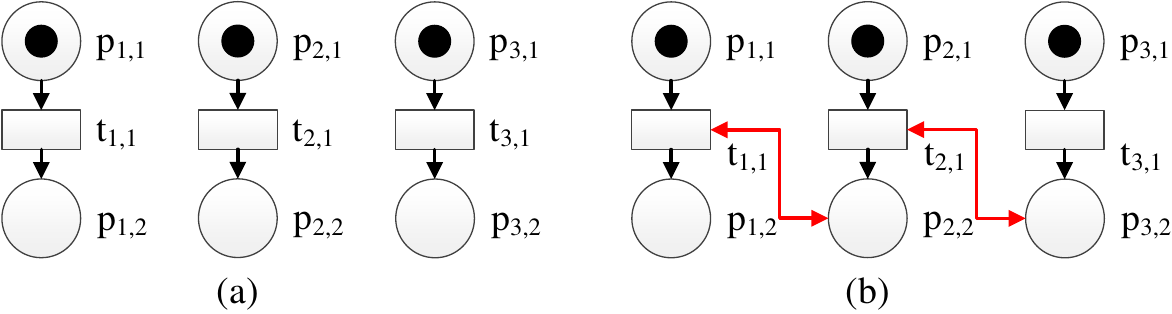}
\caption {A simple example in parallel and sequential patterns.}
\label{Para}
\end{figure}

In order to understand it better, we use a simple example to illustrate it. The KPN in Fig.~\ref{Para} (a) shows a completely concurrent system where there are three agents and there is no interaction/collaboration among them. The KPN in Fig.~\ref{Para} (b) shows the corresponding sequential pattern where the three transitions are fired only in turn.

\begin{figure*} [tt]
\centering
\includegraphics[scale=0.85]{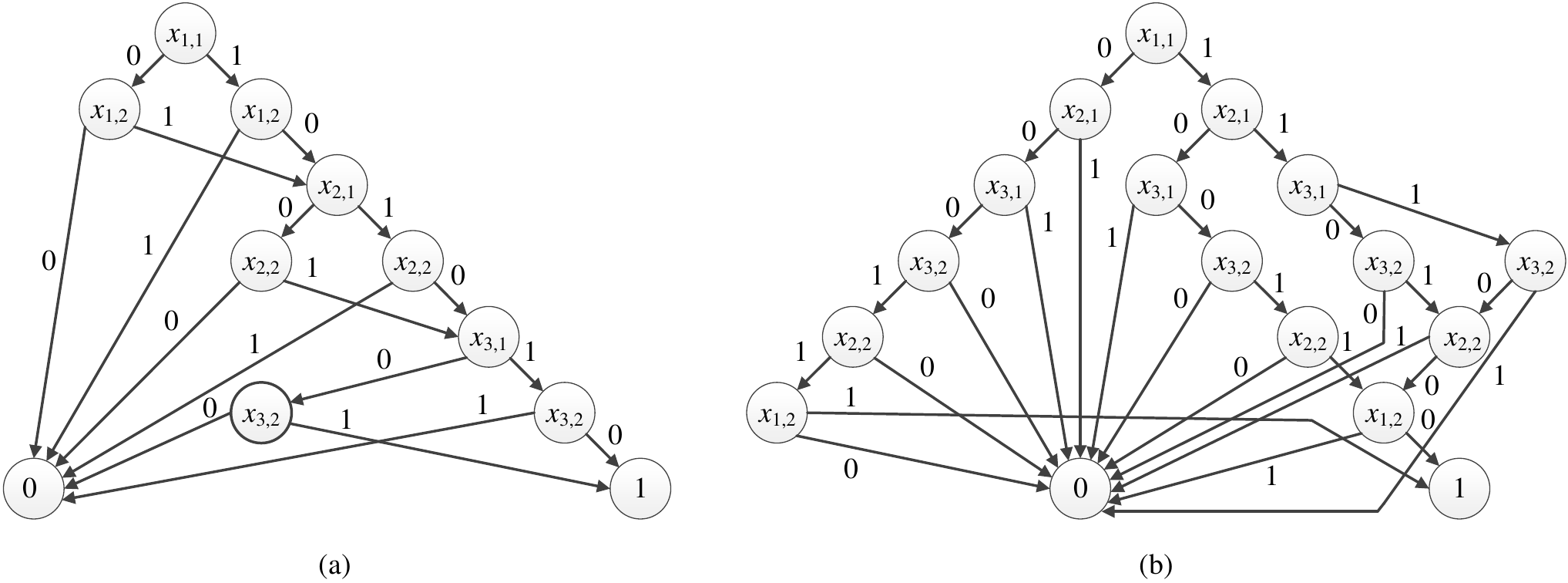}
\caption {Two OBDDs based on our heuristic method: (a) the OBDD for $f_1$; (b) the OBDD for $f_2$.}
\label{Para-OBDD}
\end{figure*}

Fig.~\ref{Para} (a) has 8 reachable markings and $\mathbb{M}$ is represented by Boolean function $f_1=(x_{p_{11}}\,\overline{x_{p_{12}}}+\overline{x_{p_{11}}}\,x_{p_{12}})\,(x_{p_{21}}\,\overline{x_{p_{22}}}+\overline{x_{p_{21}}}\,x_{p_{22}})\,(x_{p_{31}}\,\overline{x_{p_{32}}}+\overline{x_{p_{31}}}\,x_{p_{32}})$. A variable order constructed by our heuristic method is $x_{p_{11}}<x_{p_{12}}<x_{p_{21}}<x_{p_{22}}<x_{p_{31}}<x_{p_{32}}$, and the related OBDD for $f_1$ is shown in~Fig.~\ref{Para-OBDD} (a). In this and subsequent OBDDs, we use $x_i$ to represent $x_{p_i}$ for simplicity. Fig.~\ref{Para} (b) has 4 reachable markings and $\mathbb{M}$ is represented by Boolean function $f_2=x_{p_{11}}\,\overline{x_{p_{12}}}\,x_{p_{21}}\,\overline{x_{p_{22}}}(x_{p_{31}}\,\overline{x_{p_{32}}}+\overline{x_{p_{31}}}\,x_{p_{32}})+(x_{p_{11}}\,\overline{x_{p_{12}}}+\overline{x_{p_{11}}}\,x_{p_{12}})\,\overline{x_{p_{21}}}\,x_{p_{22}}\,\overline{x_{p_{31}}}\,x_{p_{32}}$. A variable order constructed by our heuristic method is $x_{p_{11}}<x_{p_{21}}<x_{p_{31}}<x_{p_{32}}<x_{p_{22}}<x_{p_{12}}$, and the related OBDD for $f_2$ is shown in~Fig.~\ref{Para-OBDD} (b). Obviously, the constructed order for the parallel pattern is better than the sequential pattern's because OBDD only uses 11 nodes to encode $f_1$ representing 8 markings but it uses 17 nodes to encode $f_2$ representing 4 markings. In fact, we can also find a good variable order for the sequential pattern if we know the sequential order of agents in advance. For example, if we know that the sequential order of agents in Fig.~\ref{Para} (b) is $a_3<a_2<a_1$, then our heuristic method chooses the marked place $p_{31}$ in the first iteration, chooses the marked place $p_{21}$ in the second iteration and chooses the marked place $p_{11}$ in the last iteration. Then the variable order is $x_{p_{31}}<x_{p_{32}}<x_{p_{21}}<x_{p_{22}}<x_{p_{11}}<x_{p_{12}}$ and the related OBDD for $f_2$ only has 13 nodes. However, it is almost impossible for a large KPN to know this sequential order so our constructed variable order for the sequential pattern is worse than the parallel pattern's.

\renewcommand\arraystretch{1.2}
\begin{table*}[tt]
\footnotesize
\caption{Experimental results of KPNer for the benchmark in parallel and sequential patterns using our heuristic method}
\label{tab-K-K}
\tabcolsep 3.65pt 
\begin{threeparttable}
\begin{tabular*}{\textwidth}{ccccccccccc}
\toprule
  \multirow{2}*{\makecell{No. of\\ cryptos\\($n$)}}&\multicolumn{5}{c}{Parallel pattern}&\multicolumn{5}{c}{Sequential pattern}\\
\cmidrule(r){2-6} \cmidrule(r){7-11}
~&$|\mathbb{M}|$&$T_{11}$ (s)&$T_{12}$ (s)&$T_{13}$ (s)&OBDD memory (B)&$|\mathbb{M}|$&$T_{11}$ (s)&$T_{12}$ (s)&$T_{13}$ (s)&OBDD memory (B)\\\hline
  $10$&$3.788\times 10^9$&$<0.001$&$0.14$&$<0.001$&$1.711\times 10^7$&$75783$&$<0.001$&$0.031$&$0.032$&$1.392\times 10^7$\\
  $20$&$3.778\times 10^{18}$&$<0.001$&$0.844$&$0.078$&$4.026\times 10^7$&$1.615\times 10^8$&$0.016$&$0.187$&$0.36$&$3.383\times 10^7$\\
  $30$&$2.964\times 10^{27}$&$0.031$&$1.922$&$0.312$&$3.917\times 10^7$&$2.513\times 10^{11}$&$0.031$&$0.562$&$1.219$&$6.048\times 10^7$\\
  $40$&$2.094\times 10^{36}$&$0.093$&$3.422$&$0.656$&$4.001\times 10^7$&$3.452\times 10^{14}$&$0.093$&$1.188$&$3.031$&$5.752\times 10^7$\\
  $100$&INF&$1.11$&$26.218$&$10.734$&$5.298\times 10^7$&INF&$1.359$&$17.547$&$113.829$&$7.406\times 10^7$\\
  $200$&INF&$8.875$&$122.625$&$154.75$&$7.741\times 10^7$&INF&$10.641$&$166.922$&$1224.36$&$2.064\times 10^8$\\
  $300$&INF&$29.64$&$299.189$&$557.843$&$1.092\times 10^8$&INF&$35.5$&$539.438$&$4465.499$&$4.213\times 10^8$\\
  $400$&INF&$69.969$&$553.625$&$1356.422$&$1.471\times 10^8$&INF&$85.687$&$1314.53$&$11326.015$&$7.714\times 10^8$\\
  $500$&INF&$136.547$&$879.219$&$2828.828$&$1.876\times 10^8$&INF&$174.797$&$3150.7$&$25823.5$&$1.109\times 10^9$\\
  $600$&INF&$238.094$&$1310.19$&$4864.157$&$2.278\times 10^8$&INF&$286.985$&$4501.11$&$42816.813$&$1.627\times 10^9$\\
  $700$&INF&$376.203$&$1727.67$&$7789.172$&$2.623\times 10^8$&INF&$456.938$&$7134.03$&Timeout&--\\
  $800$&INF&$560.063$&$2339.58$&$12051.735$&$2.966\times 10^8$&INF&$677.187$&$10664.9$&Timeout&--\\
  $900$&INF&$799.907$&$2897.12$&$16979.125$&$3.308\times 10^8$&INF&$970.609$&$15823$&Timeout&--\\
  $1000$&INF&$1098.98$&$3637.41$&$23837.75$&$3.651\times 10^8$&INF&$1375.31$&$27587.2$&Timeout&--\\
  $1100$&INF&$1455.59$&$4461.66$&$31624.157$&$4.002\times 10^8$&INF&$1909.92$&$30046$&Timeout&--\\
  $1200$&INF&$1900.86$&$5274.44$&$41985.126$&$4.355\times 10^8$&INF&$2282.92$&$36949.1$&Timeout&--\\
  $1300$&INF&$2441.74$&$6465.81$&Timeout&--&--&$3015.25$&Timeout&Timeout&--\\
\bottomrule
\end{tabular*}
\begin{itemize}[leftmargin=*]
\item The meanings of $T_{11}$, $T_{12}$ and $T_{13}$ are the same with that in TABLE~\ref{tab-K-M}.
\end{itemize}
\end{threeparttable}
\end{table*}

\subsection{Experiments of KPNer Using Two Heuristic Methods}
In fact, there has been a heuristic method~\cite{heu} to construct a static variable order in OBDD and it is also used for some Petri net analysis tools such as MARCIE~\cite{MARCIE}. Many experiments~\cite{MARCIE}
have proven that it has a good performance and thus MARCIE can obtain the first place twice in the Model Checking Contest @ Petri Nets 2015 and 2016 \cite{MCC2015,MCC2016}. Here, we compare it with our heuristic method. We first introduce it. The heuristic process is related to net structure only while ours is also related to the initial marking of a net. First, let $x_1<x_2<\cdots<x_{|P_S\cup P_K|}$, $S$ be the set of places already assigned to some variables and $S=\emptyset$ initially. Second, it computes weights $W(p)$ for all places $p\in (P_S\cup P_K)\setminus S$ according to the following formulas:

$$W(p)=\frac{f(p)}{|^\bullet p\cup p^\bullet|}$$

\begin{equation}
\begin{split}
f(p)=&\sum\limits_{t\in ^\bullet p\wedge |^\bullet t|\neq 0}\Big(\frac{g_1(t)}{|^\bullet t|}\Big)+\sum\limits_{t\in ^\bullet p\wedge |t^\bullet|\neq 0}\Big(\frac{g_2(t)}{|t^\bullet|}\Big)+\\&\sum\limits_{t\in p^\bullet\wedge |^\bullet t|\neq 0}\Big(\frac{|^\bullet t\cap S|+1}{|^\bullet t|}\Big)+\sum\limits_{t\in p^\bullet\wedge |t^\bullet|\neq 0}\Big(\frac{h(t)}{|t^\bullet|}\Big)\nonumber
\end{split}
\end{equation}

$$ g_1(t)=\left\{
\begin{aligned}
& 0.1&if\,|^\bullet t\cap S|=0\\
& |^\bullet t\cap S|&otherwise \\
\end{aligned}
\right.
$$
$$ g_2(t)=\left\{
\begin{aligned}
& 0.1&if\,|t^\bullet\cap S|=0\\
& 2\cdot|t^\bullet\cap S|&otherwise \\
\end{aligned}
\right.
$$
$$ h(t)=\left\{
\begin{aligned}
& 0.2&if\,|t^\bullet\cap S|=0 \\
& 2\cdot|t^\bullet\cap S|&otherwise \\
\end{aligned}
\right.
$$

Then the place $p$ with the highest weight $W(p)$ is assigned to the variable $x_{|P_S\cup P_K|}$ and is added to $S$. It continues the second step, and then another place with the highest weight is assigned to the variable $x_{|P_S\cup P_K|-1}$. Similarly, it repeats the second step until the last place is assigned to variable $x_1$. Then the variable order is constructed.

In our tool KPNer, we replace our heuristic method with this heuristic method and TABLE~\ref{tab-K-K-2} shows its experimental results for the benchmark in the two patterns. Obviously, this heuristic method outperforms ours a little for the sequential pattern, but is much worse than ours for the parallel pattern. The memory of our PC overflows when the number of cryptographers in the parallel pattern is 20 ($1.615\times 10^8$ states) due to the node explosion of OBDD. But our method can encode about $10^{1080}$ states of 1200 cryptographers and there is no overflow (see TABLE~\ref{tab-K-K}). We think that this heuristic method is more suitable for those KPNs with a low degree of concurrency. In this heuristic process, it also pursuits to keep dependent places as close as possible. However, it only considers net structure so that those end places in the same structure are first close to each other, which iteratively makes other places in the same structure close. Therefore, it cannot guarantee that dependent places are always as close as possible. Especially for some completely
concurrent systems, end places in the same structure are close to each other in the first iteration, and then some places in the same structure are close to each other in the second iteration. It repeats the second step until all those places in the same structure are close to each other. There are almost no dependent places close to each other in the constructed variable order. This is the reason why it performs so poorly in the parallel pattern. However, the number of these end places in the same structure is small in the sequential pattern so that it can make dependent places as close as possible in its heuristic process. This is the reason why it performs better than ours in the sequential pattern.

\begin{figure*} [tt]
\centering
\includegraphics[scale=0.77]{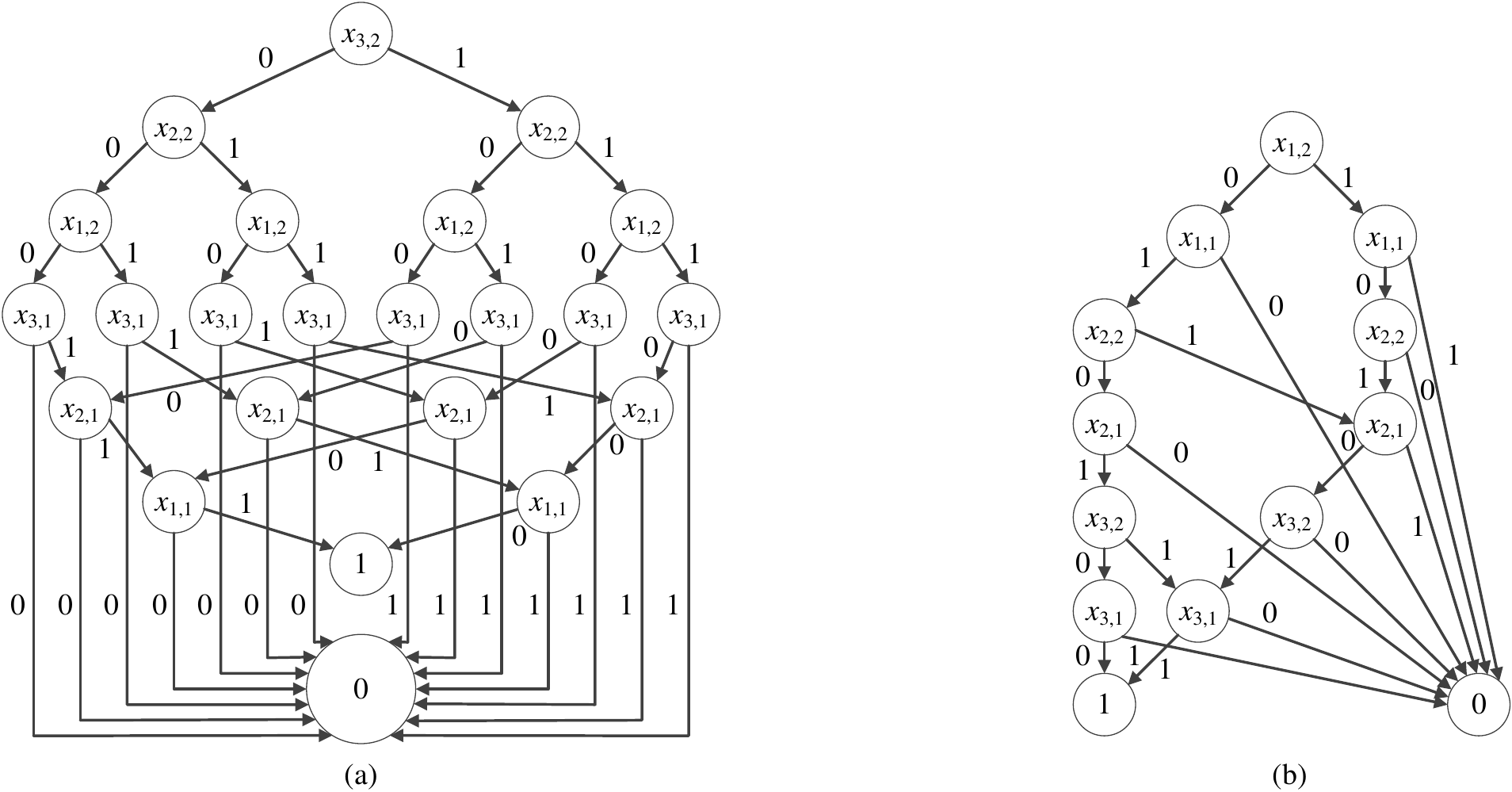}
\caption {Two OBDDs based on the heuristic method in~\cite{heu}: (a) the OBDD for $f_1$; (b) the OBDD for $f_2$.}
\label{Para-OBDD-2}
\end{figure*}

In order to understand the (dis-)advantage of this heuristic method, we also use the KPNs in Fig.~\ref{Para} to illustrate it. For Fig.~\ref{Para} (a), a variable order constructed by this heuristic method is $x_{p_{32}}<x_{p_{22}}<x_{p_{12}}<x_{p_{31}}<x_{p_{21}}<x_{p_{11}}$, and the related OBDD for $f_1$ is shown in~Fig.~\ref{Para-OBDD-2} (a). For Fig.~\ref{Para} (b), a variable order constructed by this heuristic method is $x_{p_{12}}<x_{p_{11}}<x_{p_{22}}<x_{p_{21}}<x_{p_{32}}<x_{p_{31}}$, and the related OBDD for $f_2$ is shown in~Fig.~\ref{Para-OBDD-2} (b). Obviously, the first constructed order is much worse than ours (see Fig.~\ref{Para-OBDD} (a)) but the second constructed order is a little better than ours (see Fig.~\ref{Para-OBDD} (b)). For this heuristic method, OBDD uses 23 nodes to encode $f_1$ and uses 13 nodes to encode $f_2$, while ours are 11 and 17, respectively.

\renewcommand\arraystretch{1.2}
\begin{table*}[tt]
\footnotesize
\caption{Experimental results of KPNer for the benchmark in parallel and sequential patterns using the heuristic method in~\cite{heu}}
\label{tab-K-K-2}
\tabcolsep 4.42pt 
\begin{threeparttable}
\begin{tabular*}{\textwidth}{ccccccccccc}
\toprule
  \multirow{2}*{\makecell{No. of\\ cryptos\\($n$)}}&\multicolumn{5}{c}{Parallel pattern}&\multicolumn{5}{c}{Sequential pattern}\\
\cmidrule(r){2-6} \cmidrule(r){7-11}
~&$|\mathbb{M}|$&$T_{11}$ (s)&$T_{12}$ (s)&$T_{13}$ (s)&OBDD memory (B)&$|\mathbb{M}|$&$T_{11}$ (s)&$T_{12}$ (s)&$T_{13}$ (s)&OBDD memory (B)\\\hline
  $10$&$3.788\times 10^9$&$0.015$&$3095.66$&$342.047$&$2.546\times 10^9$&$75783$&$0.031$&$0.016$&$<0.001$&$1.139\times 10^7$\\
  $20$&--&--&--&--&Overflow&$1.615\times 10^8$&$0.172$&$0.078$&$0.109$&$1.53\times 10^7$\\
  $30$&--&--&--&--&Overflow&$2.513\times 10^{11}$&$0.594$&$0.172$&$0.234$&$2.926\times 10^7$\\
  $40$&--&--&--&--&Overflow&$3.452\times 10^{14}$&$1.375$&$0.406$&$0.704$&$4.63\times 10^7$\\
  $100$&--&--&--&--&Overflow&INF&$20.062$&$4.688$&$11.687$&$6.071\times 10^7$\\
  $200$&--&--&--&--&Overflow&INF&$160.72$&$47.359$&$181.827$&$1.157\times 10^8$\\
  $300$&--&--&--&--&Overflow&INF&$548.968$&$160.719$&$958.75$&$2.367\times 10^8$\\
  $400$&--&--&--&--&Overflow&INF&$1285.81$&$374.578$&$2671.202$&$4.057\times 10^8$\\
  $500$&--&--&--&--&Overflow&INF&$2507.48$&$768.672$&$5119.187$&$6.201\times 10^8$\\
  $600$&--&--&--&--&Overflow&INF&$4298.02$&$1271.03$&$9296.39$&$8.883\times 10^8$\\
  $700$&--&--&--&--&Overflow&INF&$6879.7$&$2109.5$&$15927.813$&$1.205\times 10^9$\\
  $800$&--&--&--&--&Overflow&INF&$10295.3$&$3213.2$&$25735.781$&$1.564\times 10^9$\\
  $900$&--&--&--&--&Overflow&INF&$14915.8$&$4561.81$&$38845.391$&$1.965\times 10^9$\\
  $1000$&--&--&--&--&Overflow&INF&$20708.7$&$6481.37$&Timeout&--\\
  $1100$&--&--&--&--&Overflow&INF&$27420.8$&$8602.09$&Timeout&--\\
  $1200$&--&--&--&--&Overflow&INF&$35027.5$&$10903.1$&Timeout&--\\
  $1300$&--&--&--&--&Overflow&--&Timeout&--&--&--\\
\bottomrule
\end{tabular*}
\begin{itemize}[leftmargin=*]
\item The meanings of $T_{11}$, $T_{12}$ and $T_{13}$ are the same with that in TABLE~\ref{tab-K-M}.
\item Overflow means that the memory of our PC overflows.
\end{itemize}
\end{threeparttable}
\end{table*}

\section{Conclusion}
This paper extends the work of our conference paper~\cite{KPN}. We consider more epistemic operators that can specify more complex epistemic properties. We use OBDD to compress the state space, improve our model checking algorithms greatly and develop a tool KPNer. As shown in our experiments, KPNer is more efficient in comparison with the state-of-the-art CTLK model checker MCMAS. These advantages are owed to the combination of the OBDD technique and the structure characteristics of KPN. The combination can be seen in Algorithms~\ref{alg:order}--\ref{alg:eq} and \ref{alg:ex}--\ref{alg:c}.

The following facts should be noted. The part of the interacting/collaborating process in a KPN model is an abstraction and simulation of the execution process of an MAS, while the part of epistemic progress in the KPN possibly does not describe any related variables or actions of the MAS and some knowledge places are intentionally added into it by a designer or checker (e.g., those knowledge places in Fig.~\ref{ABP} do not correspond to any variable in such a real protocol). But the latter can indeed reflect the epistemic progresses of agents. More importantly, such a KPN with knowledge places can be used to prove whether the interacting/collaborating process of the system has some flaws which can be utilized by attackers to carry out some attacks.

Our future work includes: 1) we plan to simplify CTLK formulas and thus continually optimise our model checking algorithms; 2) we study the automatical conversion from a CTLK formula to its ENF; 3) we explore more complex epistemic operators to specify and verify more complex epistemic properties; and 4) we plan to add probability or data into KPN so that they can simulate concurrent systems more precisely and verify more design requirements.

The unfolding technique of Petri nets is also important and efficient on alleviating the state explosion problem~\cite{OBDD-SMC1,DLX2018,EV2002}, especially for Petri nets with a high concurrence degree. We plan to study the unfolding-based CTLK verification. But some difficulties should be considered, e.g., how to compute the equivalent markings of given markings through a finite complete prefix of the unfolding of a KPN?

\end{document}